\documentclass[num-refs]{nbdt-article}

\usepackage{siunitx}

\papertype{Original Article}
\paperfield{Journal Section}

\title{Comparing representational geometries using whitened unbiased-distance-matrix similarity}

\author[1,2,3]{Jörn Diedrichsen}
\author[1]{Eva Berlot}
\author[1,3,4]{Marieke Mur}
\author[5]{Heiko H. Schütt}
\author[1]{Mahdiyar Shahbazi}
\author[5,6,7]{Nikolaus Kriegeskorte}

\affil[1]{Brain and Mind Institute}
\affil[2]{Department of Statistical and Actuarial Sciences}
\affil[3]{Department of Computer Science}
\affil[4]{Department of Psychology, Western University, London, Ontario, Canada}
\affil[5]{Zuckerman Mind Brain Behavior Institute }
\affil[6]{Department of Psychology}
\affil[7]{Department of Neuroscience, Columbia University, New York, USA}

\corraddress{Jörn Diedrichsen, Brain and Mind Institute, Western University, London, Ontario, N6A3K7}
\corremail{jdiedric@uwo.ca}

\fundinginfo{National Science and Engineering Research Council: RGPIN-2016-04890; the Canada First Research Excellence Fund (BrainsCAN); German Research Foundation: DFG SCHU 3351/1-1}

\runningauthor{Diedrichsen et al.}

\begin{document}

\maketitle

\begin{abstract}
Representational similarity analysis (RSA) tests models of brain computation by investigating how neural activity patterns reflect experimental conditions. Instead of predicting activity patterns directly, the models predict the geometry of the representation, as defined by the representational dissimilarity matrix (RDM), which captures how similar or dissimilar different activity patterns associated with different experimental conditions are. RSA therefore first quantifies the representational geometry by calculating a dissimilarity measure for each pair of conditions, and then compares the estimated representational dissimilarities to those predicted by each model. Here we address two central challenges of RSA: First, dissimilarity measures such as the Euclidean, Mahalanobis, and correlation distance, are biased by measurement noise, which can lead to incorrect inferences. Unbiased dissimilarity estimates can be obtained by crossvalidation, at the price of increased variance. Second, the pairwise dissimilarity estimates are not statistically independent, and ignoring this dependency makes model comparison statistically suboptimal. We present an analytical expression for the mean and (co)variance of both biased and unbiased estimators of the squared Euclidean and Mahalanobis distance, allowing us to quantify the bias-variance trade-off. We also use the analytical expression of the covariance of the dissimilarity estimates to whiten the RDM estimation errors. This results in a new criterion for RDM similarity, the whitened unbiased RDM cosine similarity (WUC), which allows for near-optimal model selection combined with robustness to correlated measurement noise.
% Please include a maximum of seven keywords

\end{abstract}

\section{Introduction}

Systems neuroscience investigates how patterns of brain activity implement the computational processes that support behavior. The computations can be understood as transformations of representations that reflect task-relevant information about the external world, the state of the animal's body, its needs, goals, plans, or actions. Models of brain computation seek to explain how the brain processes information \cite {Kriegeskorte2019}. One way to test such models formally is to compare the representations in the models to the activity patterns measured in the brain. An essential challenge for computational neuroscience, therefore, is to develop methods for comparing representations between brains and models. 

We focus here on the approach to characterize brain representations at the level of the neural population \cite{Haxby2001,Hung2005}, which abstracts from the roles of individual neurons, and makes it easier to compare representations between brains and models \cite{kriegeskorte2008}. Brain representations are characterized by measuring patterns of activity across a brain region. Each activity pattern is associated with an experimental condition, for example the presentation of a particular sensory stimulus, and defines a point in the multivariate response space \cite{Edelman1998}. The distances among these points define the geometry of the representation. If the noise is isotropic and homoscedastic, each distance determines how well one could discriminate between two patterns. The distance matrix therefore determines the encoded information. The representational geometry additionally captures aspects of the format of the code, revealing, for example, what subset of the encoded information is amenable to linear readout \cite{Kriegeskorte2019}.

The analysis of representational geometries has come to be called representational similarity analysis (RSA, \cite{kriegeskorte2008}), and has been applied to data from invasive electrophysiology, fMRI, electroencephalography (EEG), or other methods. RSA proceeds in three steps: In the first step, the estimated activity patterns are used to compute a condition-by-condition representational dissimilarity matrix (RDM, see Figure 1). An important decision here is the choice of dissimilarity measure. Choices include the accuracy of pairwise decoders, correlation distance, or Euclidean and Mahalanobis distances \cite{Walther2016, Bobadilla-Suarez2019}. In a second step, the data and models are compared by relating the vector of upper-triangular elements of the data RDM (Figure 1) and the corresponding vectors for the model RDMs. Because the dissimilarity estimates typically lack units, models cannot predict the values of the dissimilarities directly. Instead models predict the ratios or ranks of the dissimilarities. Therefore, the off-diagonal elements are compared using cosine similarity, the Pearson correlation, or the Spearman or Kendall $\tau_a$ rank correlation \cite{nili2014}. In the third and final step, models are inferentially compared using frequentist parametric \cite{Ejaz2015} or non-parametric tests \cite{nili2014}.

\begin{figure}
\centering
\includegraphics[width=0.5\textwidth]{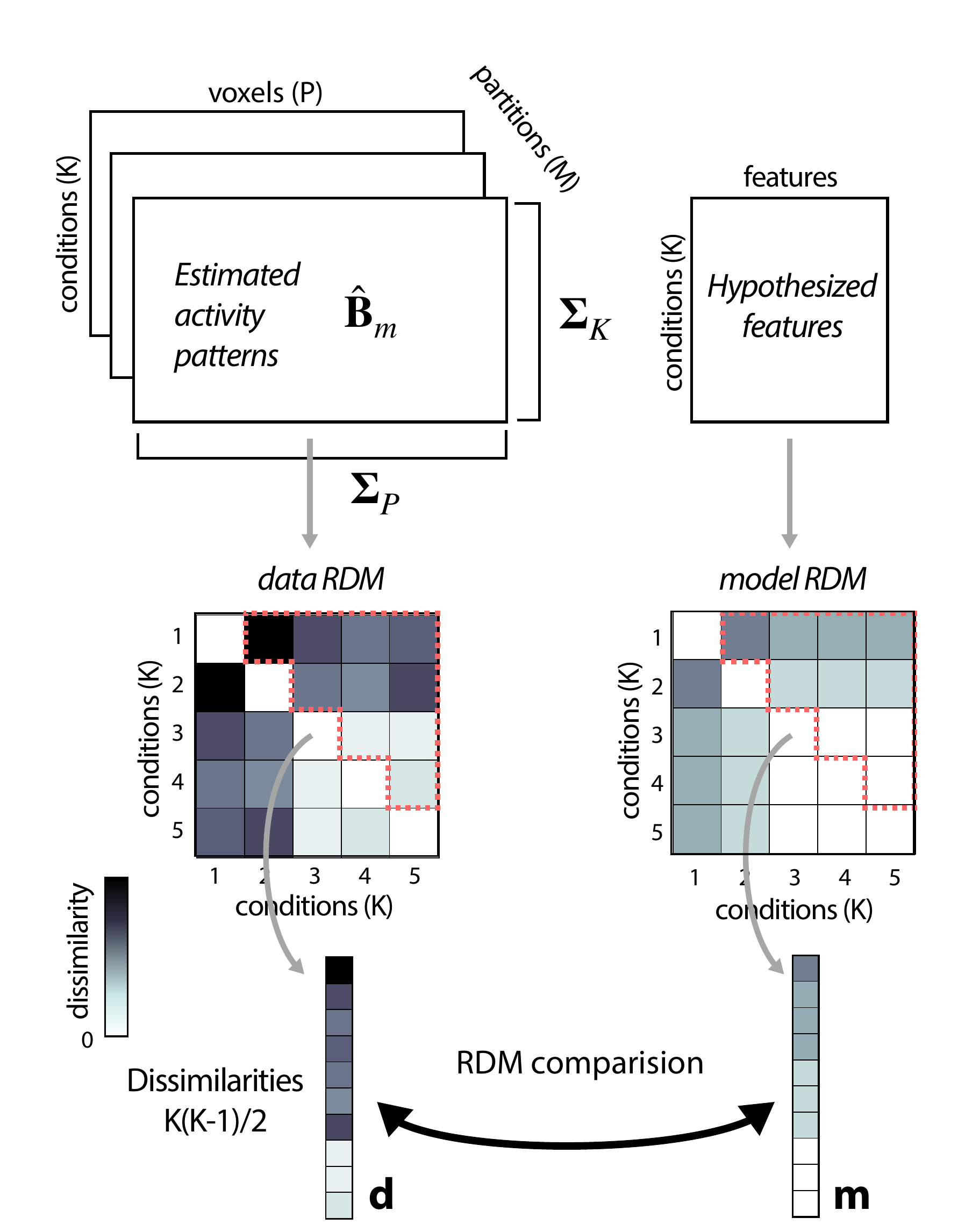}
\caption{\footnotesize{\textbf{Analysis pipeline for RSA.} The data consists of $M$ independent estimates of the activity patterns $\mathbf{B}$. A model is defined with a set of features that relate to the chosen experimental conditions and that are hypothesized to be encoded in the activity patterns. To compare data and model, the patterns are transformed into a Representational Dissimilarity Matrix (RDM). All unique pairwise dissimilarities are then stretched to a vector ($\mathbf{d}$) and compared to the vector of model dissimilarities ($\mathbf{m}$).}}
\end{figure}

In this paper, we address two closely related problems for inference using RSA. The first problem is that distances (including correlation, Euclidean, and Mahalanobis distances) are positively biased when directly estimated from noisy data \cite{Walther2016}. This can be easily seen when considering the following case: if the true activity patterns are identical, and so the true distance is zero, the measured activity patterns will still differ by virtue of the measurement noise, and the estimated distance will be larger than zero. If different conditions are measured with different noise levels, or if measurement noise is correlated across conditions, different distances will be biased to different extents. This can distort the representational geometry and potentially lead to systematic errors in the inference \cite{cai2019}. To avoid the noise-induced bias in distance estimates, we have previously proposed crossvalidated dissimilarity estimators \cite{nili2014, Walther2016}, which provide distance estimates with an interpretable 0 point. The removal of bias by crossvalidation comes at the cost of slightly increased variance. In this paper, we derive analytical expressions for the bias and variance of both biased (non-crossvalidated) and unbiased (crossvalidated) distance estimates. This allows us to gain analytic insights into when the use of unbiased distance estimates is advantageous. 

\begin{figure}
\centering
\includegraphics[width=0.8\textwidth]{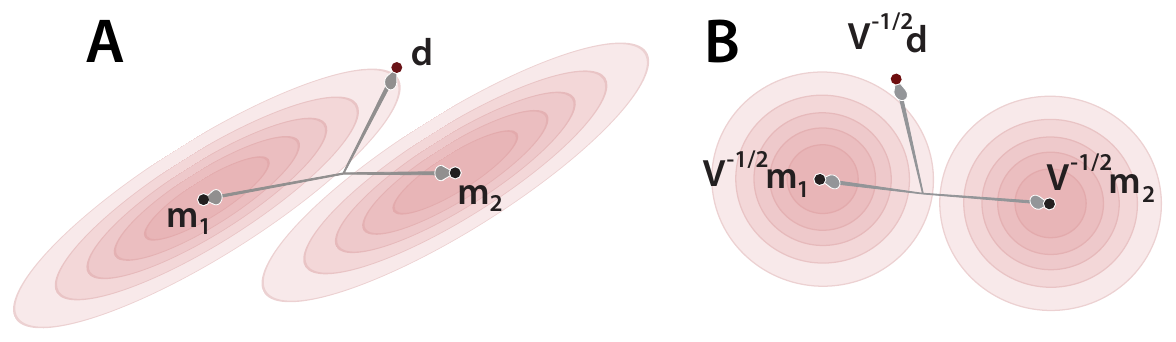}
\caption{\footnotesize{\textbf{Influence of the covariance of the dissimilarity estimates on model selection.} The data RDM $\mathbf{d}$ and two model RDMs $\mathbf{m}_1$ and $\mathbf{m}_2$ are visualized as vectors in the space spanned by the dissimilarities (one dimension for each pair of conditions). The red contours indicate the distribution of RDM estimators for different samples generated by the model. The orientation of the ellipse indicates the covariance between distance estimates. (\textbf{a}) The data RDM is closer to $\mathbf{m}_2$ in terms of the cosine similarity (angle between vectors). However, the data RDM is more likely under $\mathbf{m}_1$. (\textbf{b}) After the covariance of the distance estimates ($\mathbf{V}$) is taken into account, the data is closer to $\mathbf{m}_1$, also in terms of the angle.}}
\end{figure}

The second problem is that the elements of an estimated RDM have a complex covariance structure across different samples. This covariance, if not accounted for, can make model selection sub-optimal. The comparison between RDMs can be visualized in a space in which each unique dissimilarity of the RDM defines a dimension (Fig. 2a). When comparing RDMs with the cosine similarity, the model that has the smaller angle with the data RDM will be considered the better fit. While this approach is not systematically wrong, model comparison will not be optimal. In the second part of the paper, we therefore propose a simple method to address this issue: Using the analytical expression for the covariance of the different dissimilarity estimates, we can effectively calculate a cosine similarity in a ``whitened" space, in which the measurement error is isotropic (Fig. 2b). We show that this whitened RDM cosine similarity based on biased distance estimates is equivalent to the RV coefficient \cite{Robert1976}, which is also known as the linear Centered Kernel Alignment \cite{kornblith2019}. By combining the whitened RDM cosine similarity with unbiased distance estimates, we define a novel criterion, the \textit{whitened unbiased RDM cosine similarity} (WUC). The WUC substantially improves the power of inferential model comparisons, performing close to the theoretical maximum of the likelihood-ratio test \cite{Neyman1933, Diedrichsen2011, Diedrichsen2017Neuroimage} for normally distributed data. At the same time, the WUC is robust to violations of noise assumptions, making it the method of choice for RSA model comparison in many applications. The techniques described in this paper are all implemented in a new Python-toolbox implemented by our group \cite{Schuett2020}. 

\section{Results}
\subsection{Basic Definitions}
Let matrix $\mathbf{B}$ be the matrix of true activation values for $K$ experimental conditions measured over $P$ measurement channels. Each row of $\mathbf{B}$ contains an \textit{activity pattern} across channels, elicited by a single experimental condition. Each column of $\mathbf{B}$ contains an \textit{activity profile} across conditions, for a single channel. As common for experimental settings, we assume that the each condition is measured repeatedly, specifically that the data can be subdivided into $M$  partition, indexed by $m$, such that we have a full estimate of the activity patterns $\hat{\mathbf{B}}_m$ from each data partition. 

For example, in an fMRI experiment, the raw data are time series of blood-oxygenation-level-dependent (BOLD) signal measurements for every voxel (our measurement channels). These data is separated into $M$ different scanner runs. From these runs we can estimate, using a linear model, the activity patterns across all conditions for each run. Thus, the "measured" activity patterns would be the regression coefficients from this time series model. In other cases, such as for MEG or electrophysiological data, the activity patterns are estimated by simple averaging activity over a specific time window of recording. 

The measured activity patterns would differ slightly from run-to-run, a variation that we ascribe here to measurement noise. We consider the relatively general situation where the true signal $\mathbf{B}$ can have arbitrary structure and the measurement errors can come from any  distribution. Across the different partitions, the measured activity patterns are assumed to have covariance $\boldsymbol{\Sigma}_K$ between conditions, and a covariance $\boldsymbol{\Sigma}_P$ between measurement channels (Figure 1). Note that for the derivation of the expected variance of distance estimates, we need to make the assumption of a matrix normal distribution of the measurement noise. 

The first step in RSA is to compute the dissimilarities between activity patterns. Let $\mathbf{b}_i$ be the $i^{\rm th}$ row of $\mathbf{B}$, that is, the true activity pattern for the $i^{\rm th}$ condition across voxels. We define the $k^{\rm th}$ dissimilarity to be between conditions $i$ and $j$. In this \textit{Results} section, we consider the squared Euclidean distance, but we show in the \textit{Methods} section how the results generalize to the squared Mahalanobis distance. Both dissimilarities are based on the difference between activity patterns  $\boldsymbol{\delta}_k=\mathbf{b}_i-\mathbf{b}_j$. Specifically the squared Euclidean distance is 

\begin{equation}
d_k=(\mathbf{b}_i - \mathbf{b}_j) (\mathbf{b}_i - \mathbf{b}_j)^ T/P=\boldsymbol{\delta}_k \boldsymbol{\delta}_k ^ T/P.
\label{eq:distTrue}
\end{equation}

Note that we are normalizing all dissimilarities by the number of channels ($P$) to make the measures comparable across regions of different sizes. In an experiment with $K$ conditions, we have a total of $K(K-1)/2\equiv D$ unique pairwise distances. 

\subsection{Biased and unbiased estimates for the squared Euclidean distance}

The simplest estimator of the squared Euclidean distance can be obtained by first averaging the estimated pattern-differences across partitions

\begin{equation}
\bar{\boldsymbol{\delta}}_k=\frac{1}{M}\sum_{m}^M \hat{\boldsymbol{\delta}}_{k,m},
\label{eq:avrgPatternDiff}
\end{equation}and then taking the inner product of these estimated pattern-difference vectors. When plugging Eq. \ref{eq:avrgPatternDiff} into Eq. \ref{eq:distTrue}, we can see that the distance estimate relies on all the pairwise products of pattern differences:

\begin{equation}
\hat{d_k}=\frac{\bar{\boldsymbol{\delta}}_k \bar{\boldsymbol{\delta}}_k^T}{P} = \frac{1}{M^2}\sum_{m}^M{\sum_{n}^M{ \hat{\boldsymbol{\delta}}_{k,m}\hat{\boldsymbol{\delta}}_{k,n}^T/P}}
\label{eq:biasedDist}
\end{equation}

This estimate is positively biased by measurement noise: 

\begin{equation}
\begin{split}
\mathrm{E}(\hat{d}_k)&=\boldsymbol{\delta}_k\boldsymbol{\delta}_k^T/P+\boldsymbol{\Xi}_{kk}/M.
\end{split}
\label{eq:distbias}
\end{equation}

As shown in the methods, the expected value  ($\mathrm{E}()$) of the estimator is the true distance plus the measurement variance of the pattern difference, $\mathrm{var}(\hat{\boldsymbol{\delta}}_k)=\boldsymbol{\Xi}_{kk}$. This positive arises because we are multiplying a noisy pattern estimate with itself. 

If the measurement variance across all pattern differences is the same, the bias is a constant value across all dissimilarities, and can be accounted for by using Pearson or rank correlations to compare RDMs (see below). However, if the variance differs, the bias will systematically differ across dissimilarities, and possibly distort the representational geometry in favour of the wrong model \cite{cai2019}.

To avoid this bias, we can estimate squared distances by only multiplying pattern estimates from different, and hence independent, partitions \cite{nili2014, Walther2016} with each other. Thus, we drop from Eq. \ref{eq:biasedDist} all pairs where $m=n$. 

\begin{equation}
\tilde{d_k}=\frac{1}{M(M-1)}\sum_{m}^M \sum_{n \neq m}^M \hat{\boldsymbol{\delta}}_{k,m}\hat{\boldsymbol{\delta}}_{k,n}^T/P
\label{eq:unbiasedDist}
\end{equation}

In contrast to the biased estimate, $\hat{d_k}$, we denote the unbiased estimate as $\tilde{d_k}$. The bias is removed, as only independent partitions enter the product (for details, see Methods). Avoiding products where noise is multiplied with itself ensures that the expected value of the estimator is the distance we want to estimate: 

\begin{equation}
    \mathrm{E}(\tilde{d}_{k})=\boldsymbol{\delta}_{k}\boldsymbol{\delta}_{k}^T/P 
\end{equation}
In other words, the crossvalidated distance estimator is unbiased. 

\subsection{Variance of distance estimates}

The removal of the bias, however, does not come for free: As can be seen when comparing Eq. \ref{eq:biasedDist} and Eq. \ref{eq:unbiasedDist}, the unbiased estimate uses fewer pairs of activity patterns to estimate the true distance. We therefore expect this estimate to have a higher variance than the unbiased estimate. Indeed, as shown in the methods, we can derive a analytical expression for the variance of the biased distance if we assume normality of the measurement noise:  

\begin{eqnarray}
\mathrm{var}(\hat{d}_k)&=&\frac{1}{P^2}\left( \frac{2tr(\boldsymbol{\Sigma}_P\boldsymbol{\Sigma}_P)}{M^2}\boldsymbol{\Xi}_{kk}^2 + \frac{4P}{M}\boldsymbol{\delta}_{k}\boldsymbol{\Sigma}_P\boldsymbol{\delta}_{k}^T \boldsymbol{\Xi}_{kk} \right). \label{eq:var_biased}
\end{eqnarray} 

A very similar expression is obtained for the unbiased estimate of the distance:

\begin{eqnarray}
\mathrm{var}(\tilde{d}_k)&=&\frac{1}{P^2}\left( \frac{2tr(\boldsymbol{\Sigma}_P\boldsymbol{\Sigma}_P)}{M(M-1)}\boldsymbol{\Xi}_{kk}^2 + \frac{4P}{M}\boldsymbol{\delta}_{k}\boldsymbol{\Sigma}_P\boldsymbol{\delta}_{k}^T \boldsymbol{\Xi}_{kk} \right). \label{eq:var_unbiased}
\end{eqnarray}

Both expressions have two components: The first term of the equation arises from the multiplication of noise with noise. The variance of the distance estimate scales in the square of the measurement variance of the corresponding pattern differences ($\boldsymbol{\Xi}_{kk}$). The only difference between the biased and unbiased estimate is the size of this component, which is larger by factor $M/(M-1)$ for the unbiased estimate. The second term arises from the multiplication of the true pattern difference ($\boldsymbol{\delta}_k$) with measurement noise. If the true distance is zero, i.e. if there are no differences between the true activity patterns, this second term vanishes. The overall balance between the first and second term depends on the strength of the signal ($\boldsymbol{\delta}_k$), and on the noise covariance structure across channels ($\boldsymbol{\Sigma}_P$).

\begin{figure}
\centering
\includegraphics[width=0.7\textwidth]{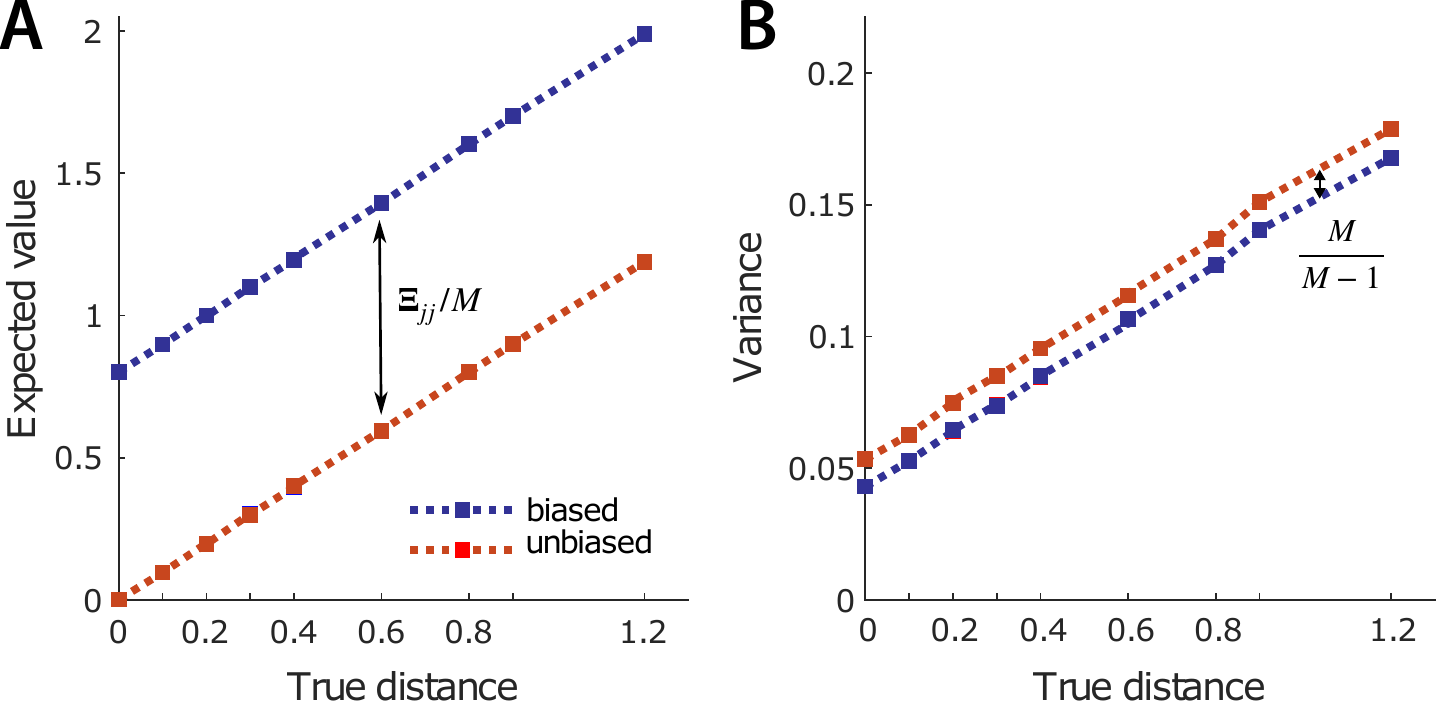}
\caption{\textbf{Bias-variance trade-off of biased and unbiased distance estimates.} (\textbf{A}) The mean of biased (blue) and unbiased (red) estimates of the squared Euclidean distance, plotted against the true value of the distances. (\textbf{B}) Variance of the distance estimates increases linearly with the true distance. The unbiased distance shows an increased variance by a factor depending on the number of partitions ($M$).}
\end{figure}

The insights from the equations are summarized in Figure 3, which shows the mean and variance of the squared distance estimates for a range of true distances between 0 and 1.2. If the true distance is 0, the mean of the unbiased estimate is zero, whereas the mean of the biased estimate is inflated by $\boldsymbol{\Xi}_{kk}/M$. In exchange, the variance of the biased estimate is lower by a factor of $(M-1)/M$. This difference is caused by using different numbers of pairwise products: The biased estimate uses all $M^2$ possible pairs, whereas the unbiased estimate excludes $M$ of the pairs (those of each partition with itself). Thus, with $M=2$ partitions, the variance of the unbiased estimate will be twice as large. However, the difference diminishes as the number of independent partitions increases. The second term of Eq. \ref{eq:var_biased}, \ref{eq:var_unbiased} causes the variance of the distance estimate to increase linearly with the true squared distance. This signal-dependence affects biased and unbiased estimates equally. 

\subsection{Model comparison using RDM correlations or cosine similarity}
Whether it is better to use biased or unbiased distance estimates depends on how these estimates are used in subsequent inference. A common use case is to compare the measured RDM to different competing models of neural representations. For this, the upper triangular part of the RDM (which is symmetric about a diagonal of zeros) is vectorized (Fig. 1).  The vector of estimated distances ($\hat{\mathbf{d}}$ or $\tilde{\mathbf{d}}$) is then compared with the vector of model-predicted distances ($\mathbf{m}$), and the model with the best correspondence is selected.

Which measure of correspondence is appropriate depends on the level at which the models are meant to make predictions. If the models predict merely the rank-ordering of the distances, a Spearman rank correlation (or when any of the models predict equal representational distances for different pairs of stimuli, Kendall's $\tau_A$ rank correlation) is appropriate \cite{nili2014}. If the models make predictions about distances on an interval scale, one can use the Pearson correlation, $r(\hat{\mathbf{d}},\mathbf{m}$). Calculating a correlation (whether Kendall, Spearman, or Pearson) allows for arbitrary scaling between observed and predicted distances, which is useful because the model-predicted dissimilarities are typically in arbitrary units and the scaling of the data depends on the signal-to-noise ratio. These correlation coefficients are also invariant to an additive constant. For biased distance estimates this is useful, as it discounts a positive bias arising from noise. Note, however, that this only works if the noise influences all conditions equally, i.e. when all distances show the same constant bias. 

In contrast, cross-validated distances estimates do not have any bias, no matter whether it is varies or is constant across conditions. The resultant distances have a ratio-scale measure, where 0 is an informative point, indicating that the two patterns only differ by measurement noise. To exploit this additional information, we can compute the correlation without subtracting the mean across distances. This quantity is the RDM cosine similarity, the cosine of the angle between the vectorized data RDM and the model RDM (Fig. 2).

\begin{equation}
\tilde{r}=\frac{\tilde{\mathbf{d}}^{T}\mathbf{m}}{\sqrt{(\tilde{\mathbf{d}}^{T}\tilde{\mathbf{d}})(\mathbf{m}^{T}\mathbf{m})}}\label{eq:cosineAlpha}
\end{equation}

In summary, for models that predict the RDM on a ratio-scale  we have two consistent choices: We can either compute the cosine similarity between the model and unbiased distance estimates (Eq. \ref{eq:cosineAlpha}, in short \textit{RDM cosine similarity}), or the Pearson correlation between the model and biased distance estimates (in short \textit{RDM correlation}).

\subsection{Using RDM cosine similarity prevents bias in model selection}
RDM correlations will only account for the biasing influence of measurement noise if the positive bias is the same across all distance estimates in the RDM - that is, when all pairwise pattern differences are measured with the same variability. If one condition has a smaller number of trials than another (e.g., after the exclusion of error trials), distances involving this condition will be systematically larger (Eq. \ref{eq:distbias}). Similarly, if the measurement errors for one pair of conditions are more correlated than for another pair (e.g. because some conditions were measured with fMRI in temporal proximity), then the variance of their pattern differences will be smaller, and the distance estimates lower. In both cases, the use of RDM correlations can bias inference towards the incorrect model \cite{cai2019}.

Consider for example the two RSA models depicted in Fig. 4a. In the first model, condition 1 and 2 belong to one category, and conditions 3 and 4 belong to another category. Within-category distances are predicted to be smaller than between-category distances. In the second model, condition 1 and 3, and condition 2 and 4 belong to the same category. We simulated pure noise data, for which the measurement noise was slightly correlated (r=0.15) across neighboring conditions (i.e. between 1-2, 2-3, and 3-4). Such correlation can for example occur in fMRI experiments, when the conditions were collected in a fixed sequence, such as in a ``traveling wave'' design, often used for perceptual fMRI experiments \cite{Kikkert2016, Wandell2007}. When we used RDM correlations to compare the data to the two models, 76.8\% of simulations were attributed to model 1 and only 23.6\% to model 2. This bias arises because the measurement noise induces a similarity structure (lower distances between neighbouring conditions) that is more similar to model 1 than model 2. When comparing models using unbiased distance estimates, the model selection bias disappeared - now exactly 50\% of the null simulations were assigned to each model. 

To study the influence of this bias on model discrimination, we simulated datasets coming from each model, while varying the number of independent partitions. We again added correlated noise to each partition, and counted how often a dataset was assigned to the correct model. Decisions using RDM correlations were on average 65.1\% correct (Fig. 4a, blue line), independent of the number of partitions (note that we left the overall amount of data the same, see methods). In contrast, decisions using RDM cosine similarity improved with increasing number of partitions (Fig. 4a, red line). For 2 partitions, decisions based on RDM cosine similarity were less often correct than those based on RDM correlations, but for more than 2 partitions, decisions based on crossvalidated distance estimates performed better. This increase arises because the variance for the unbiased estimates depends on factor $M/(M-1)$. 

When we repeated the same simulation, using independent noise for each condition, the bias disappeared, and so did the disadvantage in model selection accuracy (Fig. 4b). Now decisions based on RDM correlations were always more accurate, although the difference became smaller with an increasing number of partitions. 

\subsection{RDM cosine similarity exploits informative zero-point}
Does this mean that we should use RDM correlations with biased distances estimates when the measurement noise is i.i.d.? Not necessarily: Using the RDM cosine similarity can have advantages for some model comparisons, as it exploits the additional information inherent in the zero point. Consider the two models depicted in Fig. 4c. Both models have the same category structure, but differ in the predicted ratio of the within-category to the between-category distances. To distinguish these models, a meaningful zero point is required. When calculating the correlation between two RDMs, the mean across all distances in each RDM is first subtracted. Thus, correlations remove any constant offset of the distances in model and data - and  model selection therefore remains at chance level (0.5). Inferences can only be made using the RDM cosine similarities on unbiased distance estimates.

\begin{figure}
\centering
\includegraphics[width=\textwidth]{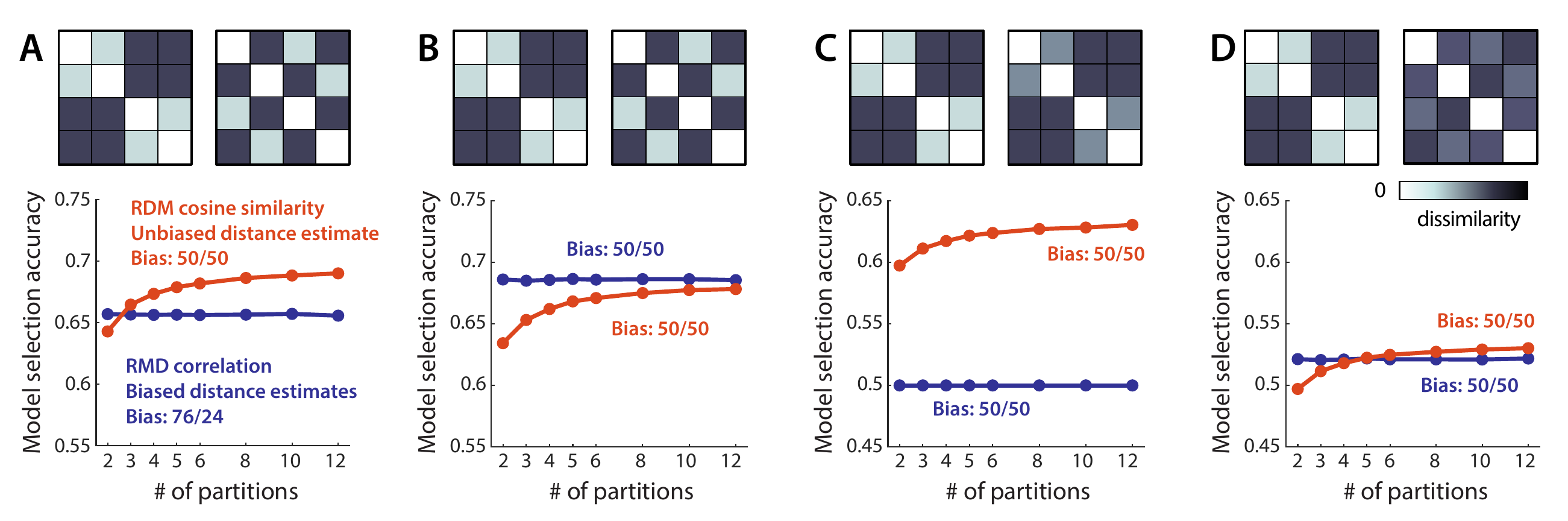}
\caption{\textbf{Model selection accuracy for RDM correlation and RDM cosine similarity.} Each of the four columns illustrates a different simulated scenario. The upper row shows the two model RDMs being compared. The lower row shows the model-selection accuracy, as a function of the number of independent crossvalidation partitions, using RDM correlation with biased distance estimators (blue) or using RDM cosine similarity using unbiased distance estimators (red). Bias: percentage of pure noise simulations for which Model1/Model2 is chosen using each criterion. (\textbf{A}) Two models with a different categorical structure for 4 conditions are compared. If the measurement noise is correlated across neighbouring conditions, model decisions using RDM correlations are biased, and perform worse than RDM cosine similarities. (\textbf{B}) The same simulation as in A, but with independent measurement noise. No bias in model decisions occurs, and RDM correlations perform better, with the advantage becoming smaller as the number of partitions increases. (\textbf{C}) When the model RDMs only differ in the ratio of the two levels of dissimilarity, RDM correlations perform at chance, because they remove the zero point which is necessary to distinguish the two models. (\textbf{D}) When the zero-point is not essential, but helps for model comparisons, using the RDM cosine similarities can be more accurate than RDM correlations even if measurement noise is i.i.d.}
\end{figure}

Finally, there are also cases in which the model decision is not fully dependent on the zero point (as in Fig. 4c), but also not fully independent of it (as in Fig. 4b). For example, the two models in in Fig. 4d have a different category structure, but also differ in the size of the distances relative to zero. As a result, we observe that RDM cosine similarities just outperform RDM correlations for larger number of partitions. Note that decisions on Null-data are unbiased in either case, as the measurement noise was modeled as i.i.d.

In summary, how the larger variability of crossvalidated distance estimates translates into model selection accuracy, depends both on the number of available partitions, as well as the structure of the two models that are being compared. In any case, unbiased distance estimates provide a safeguard against deviations from the assumption of i.i.d measurement noise --- an advantage that in many cases will be well worth the small cost in statistical power.

\subsection{Covariance of distance estimates}
The use of RDM correlations (for biased distance estimates) or RDM cosine similarity (for unbiased estimates) for model comparison would be fully adequate if all elements of the RDM were estimated independently and with the same variance. However, our analytical expression for the full covariance matrix of dissimilarities (Eq. \ref{eq:cov_biased}, \ref{eq:cov_unbiased}) shows that this is not the case, even if the underlying activity patterns are measured with i.i.d. noise.

\begin{figure}
\centering
\includegraphics[width=0.8\textwidth]{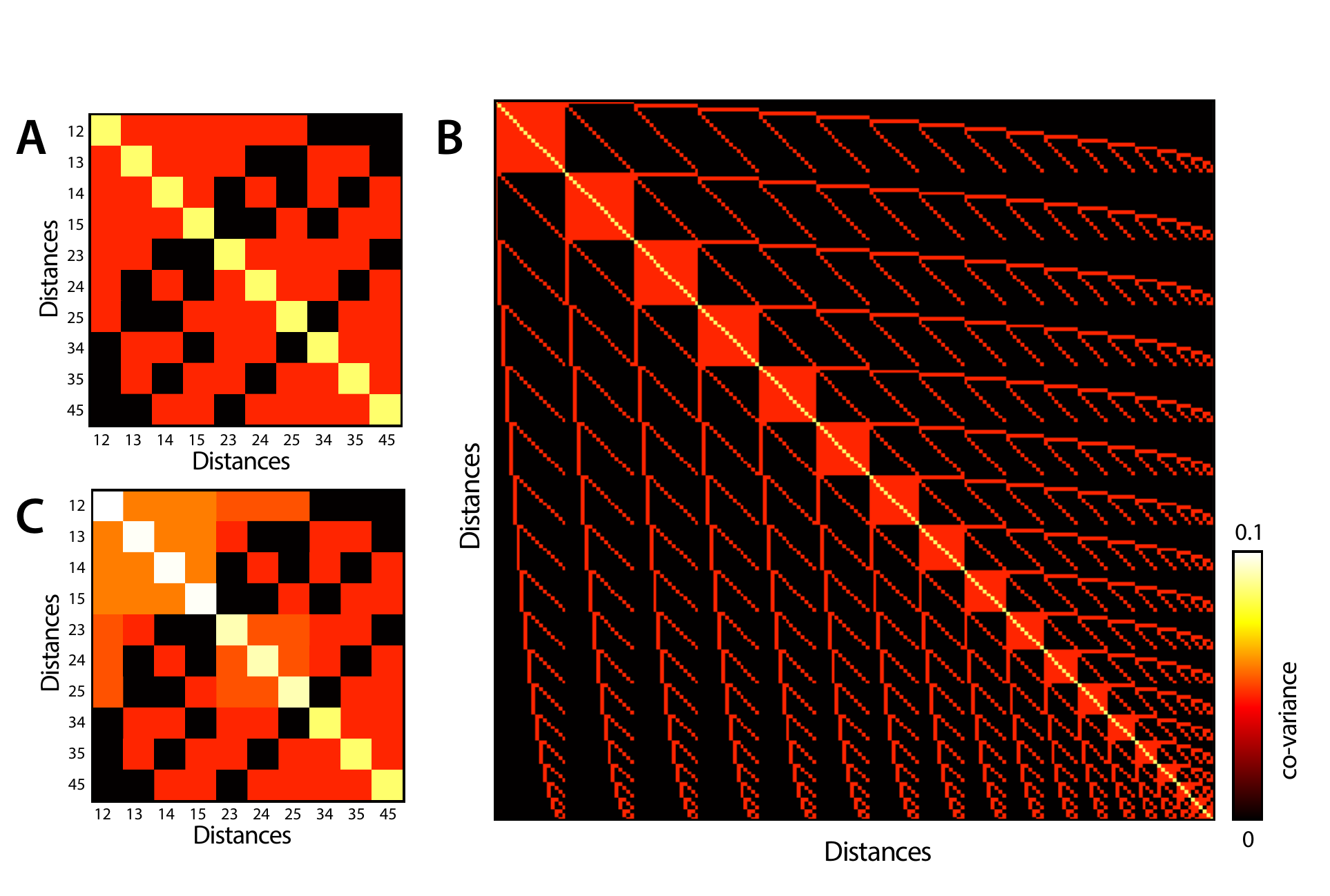}
\caption{\textbf{Covariance between elements of the RDM across independent data sets}. (\textbf{A}) Covariance matrix for the 10 distance estimates between 5 conditions, assuming that all true distances are 0 and activity patterns are measured independently and with the same variance. (\textbf{B}) Covariance matrix of the 153 distances between 18 conditions, where all the true distances are 0. (\textbf{C}) Covariance matrix for the 10 distance estimates between 5 conditions, assuming true distances as depicted in the model of Figure 1.}
\end{figure}

The correlation between distance estimates arises from the fact that, even if all conditions are measured with i.i.d. noise, the pattern difference between conditions 1 and 2 is not independent from the pattern difference between conditions 1 and 3. The covariance matrix for the 10 distance for a design with 5 conditions (assuming no true pattern differences and i.i.d. noise) is shown in Figure 5a. Distances that share one of the conditions (i.e., between $d_{1,2}$ and $d_{1,3}$) have a correlation of $r=0.25$. Only estimates of distances that do not share any conditions (i.e., $d_{1,2}$ and $d_{3,4}$) are uncorrelated. If the measurement error on the patterns is not i.i.d., a more complex co-dependence structure can arise. 

For a design with a larger number of conditions (K=18, Fig. 2b), the number of uncorrelated distances increases, i.e. the covariance matrix becomes sparser. This does not mean, however, that we can ignore the dependence structure. As we will show below, accounting for the co-dependence structure becomes especially important in designs with large numbers of conditions. 

Similar to what we have observed for the variance, the covariance between distance estimates also increases with increasing true distances. Figure 5c shows the covariance matrix for distance estimates from data generated from the RDM model depicted in Figure 1. Larger true distances exhibit larger variances and also larger covariances with other distances. The exact shape of the covariance matrix depends on the exact form of the true pattern differences $\boldsymbol{\delta}$, as well as the covariance of noise across voxels ($\boldsymbol{\Sigma}_P$).

Given the correlation between elements of the estimated RDM, we therefore would expect that both RDM correlations ($r$) and RDM cosine angles ($\tilde{r}$) will perform sub-optimally (see Figure 2). Indeed, in a previous paper \cite{Diedrichsen2017}, we have shown, for several different simulation scenarios, that ignoring this covariance structure leads to 3-12\% fewer correct model-selection decisions, as compared to a full likelihood-ratio test between models, implemented in Pattern Component Modelling \cite{Diedrichsen2011,Diedrichsen2017}. 

\subsection{RDM comparison in whitened RDM-space}
To improve model inference, we should therefore use the covariance matrix to transform the elements of the RDM into a whitened space, in which all dimensions are measured independently and with the same variance. Originally \cite{Diedrichsen2017}, we had suggested using the full expression for the covariance (Eq. \ref{eq:cov_unbiased}), and to estimate required quantities from the data. This approach, however, can be slow, as the estimation must be done iteratively, which each step involving the inversion of a $K(K-1) \times K(K-1)$ matrix - i.e. the complexity of the algorithm scales with $\mathcal{O}(K^4)$. Furthermore, for fMRI data, the spatial structure of the signal ($\boldsymbol{\delta}$) and the measurement noise ($\boldsymbol{\Sigma}_P$) are hard to estimate separately. 

We therefore propose a simplification to avoid estimating the right side of Eq. \ref{eq:cov_biased} and Eq. \ref{eq:cov_unbiased}. Specifically, we suggest using the covariance structure of the distances under the assumption that all distances are zero. In this case the variance simplifies to 

\begin{equation}
\begin{split}
\mathrm{Var}(\tilde{\mathbf{d}})&=c(\boldsymbol{\Xi}\circ \boldsymbol{\Xi}) = c \mathbf{V},
\end{split}
\end{equation}

where $\mathbf{V}$ is the structure of the covariance matrix, and $c$ is a proportionality constant that is not important for most applications. Note that noise correlations between different channels (or voxels in fMRI) would only influence $c$, but not $\mathbf{V}$ (see Discussion). The matrix  $\boldsymbol{\Xi}$ can be readily calculated from an estimate of $\boldsymbol{\Sigma}_K$ (see Eq. \ref{eq:define_Xi}). To take this covariance matrix into account for model comparison, we can prewhiten the distances by pre-multiplying them with $\mathbf{V}^{-\frac{1}{2}}$.

In the case of the cosine similarity between unbiased distance estimates, this leads to a new criterion, the whitened unbiased RDM cosine similarity (WUC): 

\begin{equation}
\tilde{r}_w=\frac{\tilde{\mathbf{d}}^{T}\mathbf{V}^{-1}\mathbf{m}}{\sqrt{(\tilde{\mathbf{d}}^{T}\mathbf{V}^{-1}\tilde{\mathbf{d}})(\mathbf{m}^{T}\mathbf{V}^{-1}\mathbf{m})}}\label{eq:whiteCosine}
\end{equation}

Similarly, we can define a whitened RDM Pearson correlation, simply by first subtracting the mean of data ($\bar{\mathbf{d}}$) and model ($\bar{\mathbf{m}}$) RDM. 

\begin{equation}
r_w=\frac{(\hat{\mathbf{d}}-\bar{\mathbf{d}})^{T}\mathbf{V}^{-1} (\mathbf{m}-\bar{\mathbf{m}})}{\sqrt{((\hat{\mathbf{d}}-\bar{\mathbf{d}})^{T}\mathbf{V}^{-1}(\hat{\mathbf{d}}-\bar{\mathbf{d}}))((\mathbf{m}-\bar{\mathbf{m}})^{T}\mathbf{V}^{-1}(\mathbf{m}-\bar{\mathbf{m}}))}}\label{eq:whiteCorrelation}
\end{equation}

\subsection{Influence of RDM whitening on model comparisons}
To determine the influence of RDM whitening in the context of realistic model comparisons, we simulated data using the designs of three published fMRI experiments and evaluated the associated models, all of which made quantitative predictions about representational distances (Fig. 6). The first two experiments measured activation patterns associated with 5 (Exp 1) or 31 (Exp 2) different finger movements in primary motor cortex \cite{Ejaz2015}. The associated models where derived either from the structure of the associated muscle activity or from the natural statistics of movement. The third experiment measured the activation patterns elicited by 92 images showing a range of animate and inanimate objects in human inferior temporal cortex. The models were derived from the 8 layers of a neural network \cite{Khaligh-Razavi2014}. 

As for Figure 4, we simulated data from each model, using different signal-to-noise levels. We then used RDM correlations or whitened RDM correlations, as well as cosine similarity or whitened cosine similarity to find the best model. For each method we recorded the number of correct model decisions. In addition to the RSA-based methods, we also used Pattern Component Modeling (PCM) \cite{Diedrichsen2011,Diedrichsen2017}, which directly compares the marginal likelihood of the data given the models, under the assumption that both signal and noise are normally distributed. For such data, PCM implements the likelihood-ratio test between models. For the case of our simulations, where all assumptions hold, PCM therefore implements the optimal inference procedure \cite{Neyman1933} and provides an upper performance bound for any model-comparison technique. 

\begin{figure}
\centering
\includegraphics[width=\textwidth]{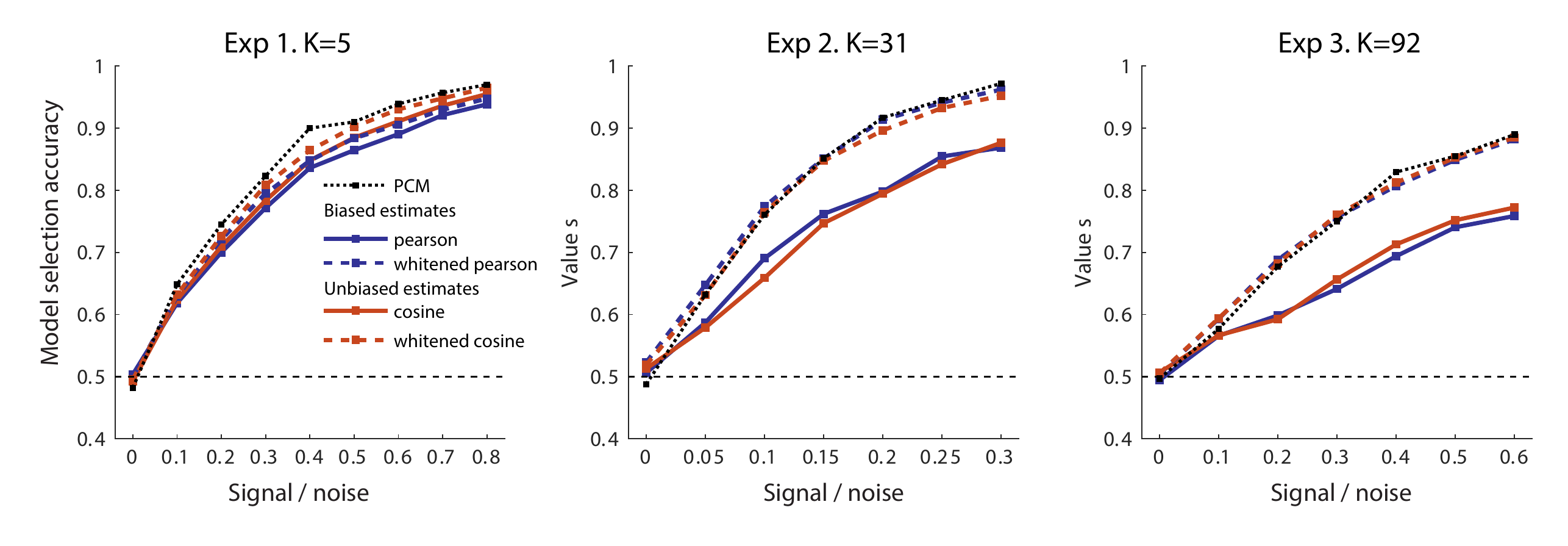}
\caption{\textbf{Model selection accuracy for different representational distance estimators and measures of RDM similarity.} A theoretical upper bound on model selection accuracy is provided by PCM (black dotted line). Model decisions are either based on RDM correlations with  biased distance estimators (blue) or on RDM cosine angles with unbiased (crossvalidated) distance estimators (red). Both RDM correlations and RDM cosine similarity can be performed in original (solid line) or in whitened space (dashed line). Different panels show simulations for three experimental designs with different numbers of experimental conditions $K$ and different models (see methods).}
\end{figure}

Across the three different experimental simulations, the simple RDM correlation or RDM cosine similarity (Fig. 6, solid lines) performed clearly sub-optimally as compared to PCM. Taking the covariance structure of the distances into account (dashed lines) substantially improved model decisions. Indeed, in many cases, the performance of RSA inference was close to optimal. This suggests that accounting for the signal-dependent, second half of the covariance formula (Eq. \ref{eq:cov_biased}, \ref{eq:cov_unbiased}) would not improve inference much further. Instead, these simulations indicate that the observed sub-optimal performance was mostly caused by the assumption that the distances are uncorrelated, rather than by the assumption that the distances have the same variance.

\subsection{Factors influencing the advantage of RDM whitening}
From the simulations in Figure 6, it appears that the importance of taking the covariance structure of the distances estimates into account is more pronounced for experiments with more conditions. To test this idea directly, we simulated data for Exp 1 with different numbers of conditions. We started with 5 conditions and 32 partitions. We then reanalyzed the data, relabeling the measures from even partitions as conditions 1-5, and from the odd partitions as condition 6-10. This increases the number of conditions to 10 and reduces the number of partitions to 16. We repeated this procedure two times more, finally ending up with 40 conditions and 4 partitions. As the underlying data is the same, the performance of PCM was relatively constant across these situations. With increasing number of conditions, however, the advantage of using WUC over the normal RDM cosine similarity increases (Fig. 7).  
This may appear at first somewhat counter-intuitive, as the proportion of uncorrelated distances pairs in the RDM increases with increasing number of conditions. However, the structure of the covariance matrix (Fig. 5b) is such that the anisotropy increases with the number of conditions. The axis of highest variability of the distance estimates is always in the direction of the average of all distances. This direction is associated with an eigenvalue of $K$. There are also $K-1$ orthogonal directions with an eigenvalue of $K/2$, and $K(K-3)$ directions with an eigenvalue of 1. Thus, the ratio of the larger to the smaller eigenvalues of $\mathbf{V}$ (a measure of anisotropy) scales linearly with the number of conditions. That means that with 40 conditions, the all-mean dimension has 40 times higher variability than most of the other directions. When ignoring the covariance structure, all dimensions are counted to be equally important, which leads to sub-optimal inferences.    
    
\begin{figure}
\centering
\includegraphics[width=0.5\textwidth]{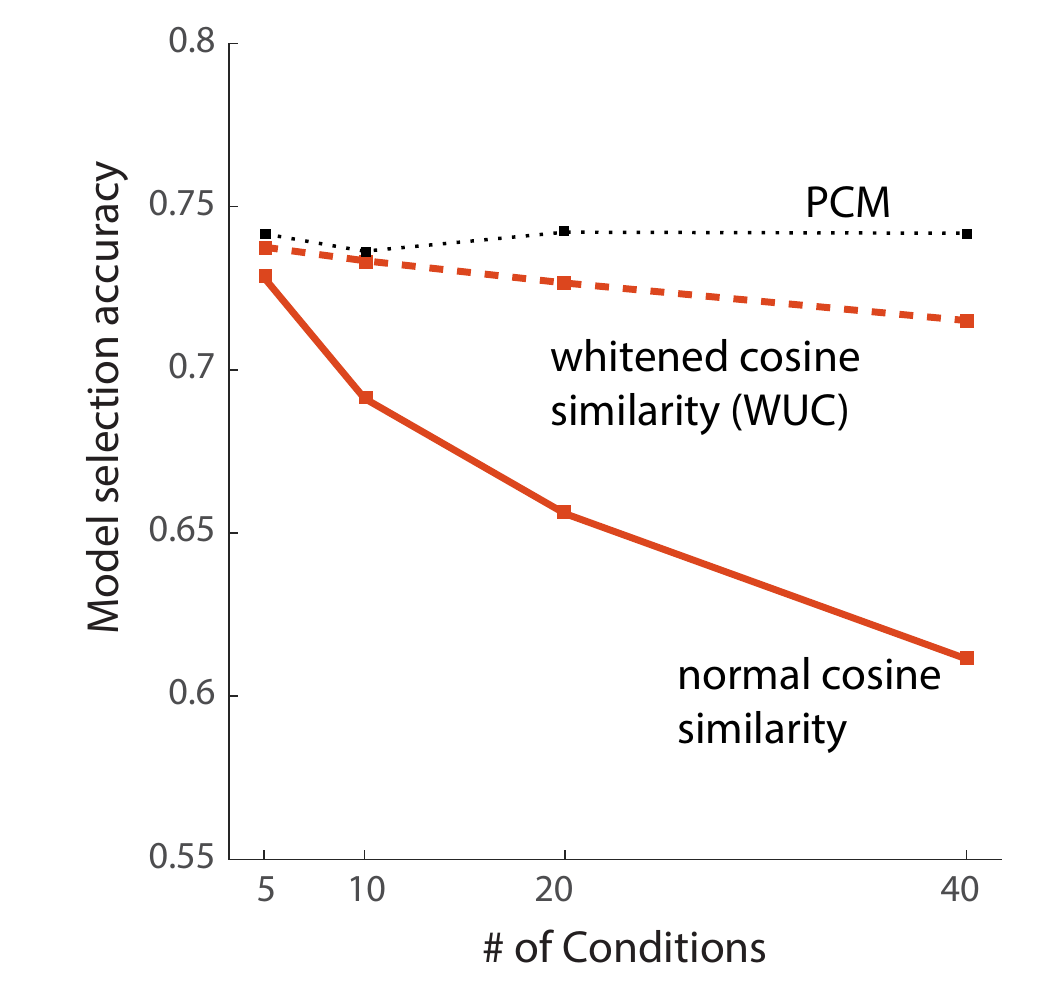}
\caption{\textbf{Model selection accuracy for normal and whitened RDM cosine similarity as a measure of RDM fit.} Model selection accuracy is greater for the whitened unbiased cosine similarity (WUC) than for the normal RDM cosine similarity based on unbiased distance estimates. The benefit of whitening grows with the number of experimental conditions. Dashed line indicates performance of likelihood-ratio test as implemented in Pattern Component Modeling (PCM).}
\end{figure}

This consideration also implies that for some model comparison problems, taking into account the covariance will not change the inference. This is the case when two models differ only on dimensions of model space that can be measured with equal variability (i.e., have the same eigenvalue in $\mathbf{V}$). The two models compared in Fig. 4a are such an example. Most model comparison problems, however, will improve, and inference will never get worse. Thus, using whitened RDM similarities is always recommended.

\subsection{Interpretation of the WUC}
An important consideration in the interpretation of the WUC is that its expected value will be zero if and only if all distances that are systematically larger than zero in the data are predicted to have a distance of zero in the model. That is, a WUC of zero shows that there is no above-chance linear prediction possible from the features of the model to the activity patterns (see Fig. 1). 

Therefore, any model that predicts a positive distance for any pair of conditions that are distinct in the data, will have a WUC $>0$. In contrast, the traditional RDM Pearson correlation will be zero when there is no systematic relationship between measured and predicted dissimilarities. As a consequence, the values of the WUC tend to be substantially higher than RDM correlations, often close to 1 for most models. 

To obtain a baseline that would be equivalent to a correlation of zero, it is therefore often useful to compare all models to a null-model that predicts all conditions as equally distinct from each other. This approach combines the interpretability of the RDM correlation with the increased power of the WUC. 

\subsection{Extension of the WUC to flexible RDM models}
In many applications, we want to fit and evaluate flexible models, in which  the vector of predicted dissimilarities depends on a vector of model parameters $\boldsymbol{\theta}$. In some cases the predicted RDM is a weighted sum of different model components, for example the different layers of a deep neural network \cite{Khaligh-Razavi2014,Khaligh2017}. In other cases, the predicted distances are a non-linear function of the model parameters \cite{Diedrichsen2017}, for example when parameterizing the width of a tuning function in a population-receptive-field model \cite{Dumoulin2016}. In all of these cases, we can take into account the covariance of the dissimilarity estimates by minimizing the following loss function:

\begin{equation}
J(\boldsymbol{\theta})=(\mathbf{d}-\mathbf{m(\boldsymbol{\theta})})^{T}\mathbf{V}^{-1}(\mathbf{d}-\mathbf{m(\boldsymbol{\theta})}).
\end{equation}

Equivalently, we can whiten the estimation error by pre-multiplying both estimated distances and the model prediction with $\mathbf{V}^{-\frac{1}{2}}$, and then use standard least squares approaches.

\subsection{Relationship to other multivariate dependence measures}
Interestingly, the whitened RDM cosine similarity is related to other statistical measures of multivariate dependence. If a whitened RDM cosine similarity is calculated from biased distance estimates, and if we assume i.i.d. measurement noise, it is identical linear Centered Kernel Alignment (CKA, \cite{kornblith2019, Cristianini2006}), also known as the RV coefficient \cite{Robert1976}. The CKA is a normalized version of the Hilbert-Schmidt independence criterion (HSIC \cite{Gretton2005}) between two sets of multivariate patterns. In general, the HSIC is the dot-product between the elements of two similarity (or kernel) matrices. Here we have the special linear case, where the similarity is the dot product of the two pattern vectors. 

Let $\mathbf{A}$ and $\mathbf{B}$ be two matrices, with the same number $K$ of rows, containing patterns for $K$ observations (e.g., trials or time points). To make the mean of each column of these matrices equal to zero, we can pre-multiply the patterns with the centering matrix $\mathbf{H} = \mathbf{I}_K - \mathbf{1}_K/K$. We then can define the centred second moment matrix of the patterns as

\begin{equation}
    \mathbf{G_A} = \mathbf{H} \mathbf{A} \mathbf{A}^T \mathbf{H}^T/P \qquad \mathbf{G_B} = \mathbf{H} \mathbf{B} \mathbf{B}^T \mathbf{H}^T/P.
\end{equation}

For the linear case, the HSIC is the dot product between the elements of the two second-moment matrices. 

\begin{equation}
\mathrm{HSIC_{A,B}} = \mathrm{vec}(\mathbf{G_A})^T \mathrm{vec}(\mathbf{G_B}) 
\label{eq:HSIC}
\end{equation}

The linear CKA is the normalized version of this quantity, just like the correlation coefficient is a normalized version of the covariance.

\begin{equation}
\mathrm{CKA} = \frac{\mathrm{HSIC}_{A,B}}{\sqrt{\mathrm{HSIC}_{A,A}\mathrm{HSIC}_{B,B}}} \label{eq:CKA}
\end{equation}

To show the equivalence of Eq. \ref{eq:whiteCosine} (using biased distance estimates) and Eq. \ref{eq:CKA}, we can express the vector of squared distances $\mathbf{d_A}$ as a linear combination of the elements of the second moment matrix,
\begin{eqnarray}
    d_{j} &=& (\mathbf{b}_i - \mathbf{b}_k)(\mathbf{b}_i - \mathbf{b}_k)^T\\
    &=& \mathbf{G}_{i,i}+\mathbf{G}_{k,k}-\mathbf{G}_{i,k}-\mathbf{G}_{k,i},
\end{eqnarray}

which we can write more succinctly using a properly defined linear transformation matrix $\mathbf{T}_d$, such that

\begin{equation}
\mathbf{d_A}=\mathbf{T}_d \mathrm{vec}(\mathbf{G_A}).
\end{equation}

The structure of the $\mathbf{T}_d \mathbf{T}_d^T$ can be shown to be identical to that of the covariance matrix of the dissimilarities (Eq. \ref{eq:cov_biased}, \ref{eq:cov_unbiased}), which is determined by the $D \times K$ contrast matrix $\mathbf{C}$ (see methods): 

\begin{equation}
\begin{split}
\mathrm{vec}(\mathbf{G_A})^T \mathrm{vec}(\mathbf{G_B})&=\mathbf{d_A}^T (\mathbf{T}_d \mathbf{T}_d^T)^{-1}\mathbf{d_B}\\
&=\mathbf{d_A}^T (\mathbf{CC}^T \circ \mathbf{CC}^T)^{-1} \mathbf{d_B}\\
&=\mathbf{d_A}^T (c\mathbf{V})^{-1} \mathbf{d_B}.
\label{eq:newCKA}
\end{split}
\end{equation}

A more intuitive explanation of this equivalence is that all unique elements of an estimate of $\mathbf{G}$ are mutually uncorrelated, if the true $\mathbf{G}=\mathbf{0}$ and $\boldsymbol{\Sigma}_K = \mathbf{I}\sigma^2$ (Eq. \ref{eq:normalExp}). The covariance-structure of the distances is then simply induced by the fact that some of the distances share common elements of the second moment matrix, with the covariance structure determined (up to a constant) by $\mathbf{T}_d  \mathbf{T}_d^T$.

While the whitened RDM cosine similarity using biased distance estimates is equivalent to the linear CKA or the RV coefficient, the whitened unbiased RDM cosine similarity (Eq. \ref{eq:unbiasedDist}, WUC) defines the unbiased version of this coefficient. For the special case of i.i.d. noise ($\boldsymbol{\Sigma}_K = \mathbf{I}\sigma^2$), it can be computed based on Eq. \ref{eq:HSIC} and \ref{eq:CKA}, replacing $\mathbf{G}$ with an unbiased estimate for the second moment matrix.

\begin{equation}
    \tilde{\mathbf{G}} = \frac{1}{M(M-1)} \sum_{m}\sum_{n\neq m}{ \mathbf{H} \hat{\mathbf{B}}_m \hat{\mathbf{B}}_n^T \mathbf{H}^T}/P
        \label{eq:unbiasedG}
\end{equation}

For the more general case of arbitrary heteroskedasticity ($\boldsymbol{\Sigma}_K \neq \mathbf{I}\sigma^2$), the WUC provides a way of incorporating a known noise covariance of the data for optimal inference. In the quadratic form in Eq. \ref{eq:newCKA}, the term $\mathbf{CC}^T \circ \mathbf{CC}^T$ can be replaced with $\mathbf{C}\boldsymbol{\Sigma}_K\mathbf{C}^T \circ \mathbf{C}\boldsymbol{\Sigma}_K\mathbf{C}^T$, ensuring that uneven measurement noise across conditions is being taken into account during model comparison.

\section{Discussion}
RSA provides an intuitive and flexible way of performing inference on representational models (i.e., on models that describe the relationship between high-dimensional activity patterns). There are, however, numerous different dissimilarity measures and ways of comparing measured RDMs to model RDMs, and the optimal way of implementing RSA remains a matter of debate \cite{nili2014, Diedrichsen2017}. In this paper, we derive an analytical expression for the mean and the covariance of the biased and unbiased estimates of squared Euclidean and Mahalanobis distances. In the following we will discuss the theoretical and practical aspects arising from these results.  

\subsection{Dealing with bias and covariance of distance estimators}
First, we show that standard distance estimates are positively biased, and that this bias depends on the variances and covariances of the measured activity patterns ($\boldsymbol{\Sigma}_K$). If the measurement noise is i.i.d. across trials, the bias will be the same across all distance estimates, and can be taken into account by ignoring the mean distance in subsequent model comparisons (for example by using the Pearson or rank correlation). If, however, one condition is measured with higher variance (for example because there were different numbers of repetitions or error trials had to be discarded), then all distances involving this condition will tend to be higher. If two conditions systematically follow each other, such that they are measured with a positive covariance, their dissimilarity will be systematically lower than two conditions measured with independent noise. These biases can translate into biases in model selection \cite{cai2019}. To avoid such errors, we can remove the bias in the estimation of the distance using crossvalidation. This approach has the substantial advantage that unequal measurement errors across conditions can no longer bias model decisions. However, removing the bias of the distance estimates comes at a cost: the variance of unbiased distance estimate is slightly higher than the variance of biased distance estimates, to be precise, by factor $M/(M-1)$. Thus, when using unbiased estimates, a large number of independent partitions ($M$) is desirable.

Second, we show that dissimilarity estimates within an RDM are systematically correlated with each other. In a previous paper, we had shown that model selection using the RDM cosine similarity or RDM correlation was less accurate than PCM \cite{Diedrichsen2017}. Here we show that taking the covariance structure of the dissimilarity estimates into account improves our power to adjudicate between models. This improvement can even be achieved when using the covariance structure predicted under the assumption that all true distances are zero, which dramatically simplifies the procedure, avoiding iterative calculation of the covariance matrix. The power achieved with the whitened RDM cosine similarity (Eq. \ref{eq:whiteCosine}) or the whitened RDM Pearson correlation (Eq. \ref{eq:whiteCorrelation}) is close to the theoretical optimum, as achieved with the likelihood-ratio test implemented by PCM. 

Taken together, these two insights suggest the use of unbiased distance estimates combined with the whitened cosine similarity to compare RDMs. We call this new approach the whitened unbiased RDM cosine similarity (WUC). It has important connections to the linear CKA \cite{kornblith2019, Cristianini2006}) and RV coefficient \cite{Robert1976},  but extends these two traditional approaches by removing the biasing influence of measurement noise by using a crossvalidated estimate of the distances (Eq. \ref{eq:unbiasedDist}) or second moment matrix (Eq. \ref{eq:unbiasedG}). 

\subsection{When should the whitened unbiased RDM cosine similarity be used? }
The optimal method of course always depends on the data and models that need to be compared  (Figure 8). The first decision is whether the models are meant to predict the dissimilarities quantitatively (ratio scale) or only their ranks (ordinal scale). Quantitative predictions can often be derived if we have an explicit model of the shape of the underlying activity profiles. The distribution of activity profiles may also be predicted from activities in an artificial neural network model \cite{Khaligh-Razavi2014}, directly from perceptual judgements \cite{Sormaz2016}, or the statistics of external training data \cite{Ejaz2015}. In other cases, the model may only predict the rank ordering of the dissimilarities, but not by how much one dissimilarity is larger than another. In such cases, rank-correlations are most appropriate \cite{nili2014}. While this approach can be statistically less powerful \cite{Diedrichsen2017}, it is robust against any possible monotonic transformation of the dissimilarities.

The next decision is whether the activity patterns can be estimated independently and with approximately the same variance across all partitions. If we cannot be sure that this is the case, then crossvalidated, unbiased distance estimates should be used. This is important, because the bias on the standard Euclidean or Mahalanobis distances will be structured if the noise is not i.i.d., such that the model comparison will be biased. Even in situations in which the measured activity-pattern estimates can be assumed to be i.i.d., the unbiased estimation approach can be more powerful than using the biased estimates and Pearson's correlation. This is because the meaningful zero point (which indicates that there is no pattern difference) can help distinguish models. Which approach is better depends on the number of partitions, the signal-to-noise ratio, the experimental conditions, and the structure of the models (Fig. 4). Overall, however, the increased robustness of violations of noise assumptions will generally outweigh the cost of increased variance, especially if the number of partitions is large. 

Whether biased or unbiased distance estimates are used, RDMs should always be compared using whitened RDM correlations or cosine similarities. These measures perform often substantially better, but never worse than standard approaches. Because we can approximate the true covariance structure well using the covariance structure under the assumption that there is no true signal, the approach can be implemented in a computationally efficient manner. Thus, we do not see any reason to use standard Pearson correlation or cosine similarity for RDM comparison.

\begin{figure}
\centering
\includegraphics[width=0.7\textwidth]{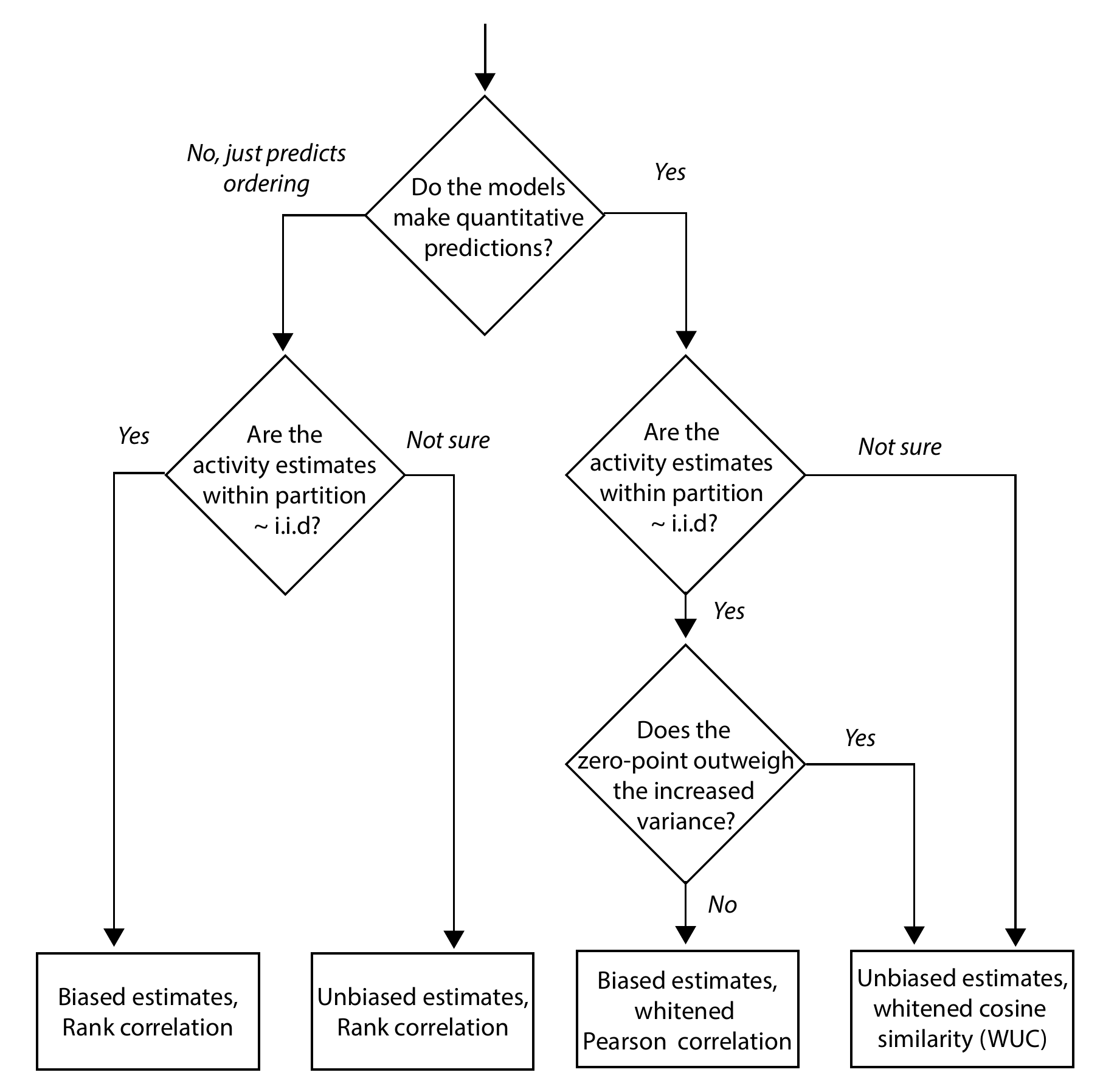}
\caption{\textbf{Decision tree for the selection of a dissimilarity measure and model comparison approach for RSA.} Rank correlations (Kendall's tau II or Spearman's correlation) are most appropriate if the models do not make quantitative predictions about the size of the distances. Unbiased dissimilarity estimates should always be used, if the activity estimates within a data partition are not measured independently and with the equal variance. Model comparisons should always be conducted in a whitened RDM space.}
\end{figure}

\subsection{Probabilistic versions of RSA}
An alternative to using unbiased distance estimates is to rely on an explicit model of the measurement noise. This approach is used in PCM \cite{Diedrichsen2017Neuroimage, Diedrichsen2011}, and two other related Bayesian formulations of RSA \cite{Friston2019, cai2019}. All three approaches rely on estimating the likelihood of the data under the RDM model, using the assumption of Gaussian signal and noise. While this method can be used to derive maximum-likelihood estimates of the distances \cite{cai2019}, these are by construction not unbiased. Rather, all three approaches rely on a valid estimate of the trial-by-trial  ($\boldsymbol{\Sigma}_K$) and spatial ($\boldsymbol{\Sigma}_P$) noise covariance for valid model inference. Here we show that, if these estimates are valid, PCM provides a theoretical upper bound for the performance of any RSA method. However, if the trial-by-trial noise covariance is not specified correctly (as in the example of Fig. 4a), model selection bias would arise, similar to the one shown for biased RDM correlations (see Fig 4a). In cases in which there is a high uncertainty about ($\boldsymbol{\Sigma}_K$), the use of WUC may therefore be preferable to PCM. 

\subsection{Influence of spatial noise correlations}
While we have extensively discussed the influence of the trial-by-trial covariance ($\boldsymbol{\Sigma}_K$) on the distance estimates, many neuroscientific datasets also have substantial spatial noise correlations between channels ($\boldsymbol{\Sigma}_P$). This is especially true for fMRI data, where strong correlations between voxels exist. For this reason, it is common practice to attenuate these correlations by spatial pre-whitening activity estimates before multivariate analysis \cite{Walther2016}. However, given that our estimate of the spatial noise structure is not perfect and often needs to be regularized, we will be left with some residual noise correlations between voxels. 

Eq. \ref{eq:cov_unbiased} shows that such correlations simply inflate the (co-)variance of the distance estimates under the null-hypothesis by a constant factor, which is related to the "effective" number of voxels, $\mathrm{tr}(\boldsymbol{\Sigma}_P)^2/\mathrm{tr}(\boldsymbol{\Sigma}_P\boldsymbol{\Sigma}_P)$. Because subsequent RSA inference, for example a t-test on distance estimates across subjects, takes the variability of distances into account directly, the inflation does not affect the validity of such inferences. In Appendix 6.3, we provide a simulation with spatially correlated noise. The results confirm the advantage of using whitened RDM correlations or RDM cosine similarities generalizes to situations with substantial noise correlations between channels. 

In contrast, the likelihood estimates from PCM directly rely on valid spatial covariance estimates. This is not a problem if we simply use the these estimates for model comparison using frequentist tests \cite{Diedrichsen2017Neuroimage}, as again, the increased variability of the likelihood estimates will be directly taken into account. However, a direct interpretation of the likelihoods, for example for a subsequent Bayesian group analysis \cite{Stephan2009}, needs to be done with some caution. 

Interestingly, the simulation (see 6.3) also shows that for very high spatial noise correlations, model comparison using WUC is slightly superior to model comparison using PCM. Thus, the likelihood-based test of PCM outperforms RSA only if the spatial noise structure is relatively close to the assumption of independence.   

\subsection{Electrophysiological and other non-Gaussian data} 
 While the derivation of the bias (Eq. \ref{eq:distbias}) and the unbiasedness of the crossvalidated distance estimator (Eq. \ref{eq:distbias2}) holds for any data distribution, the derivation of the covariance matrix of distances (Eq. \ref{eq:cov_unbiased}) depends on the assumption of a Gaussian distribution of the noise. 
 
 One important application, for which this assumption is clearly violated, are neural spiking data obtained from single- or multi-unit electrophysiological recordings. Estimates of instantaneous firing rates show a Poisson-like distribution across repeated measurements, with increasing firing variability for higher rates and a rightward skew. For such data, we recommend to take the square-root of the instantaneous firing rates before calculating distances \cite{Yu2009}. The square-root transform makes the distribution more symmetric the variance approximately constant across firing rates. 
 
To test the approach suggested here against modest violations of the Gaussian noise assumption, we repeated the simulations shown in Fig. 7 using noise distributions with either a modest skew ($\chi^2$-distribution with df=6) or a modest kurtosis (t-distribution with df=6). Overall, we were able to replicate the increasing advantage of the advantage of the WUC with increasing number of conditions. Thus, model comparison using WUC appears to be robust against modest violations of the Gaussian noise assumption, as we would find in real fMRI or electrophysiological data. 
 
\subsection{Summary}
Taken together, we believe that the WUC provides an important new measure for RDM comparison that should become standard for many applications. The new measure extends the linear centered kernel alignment (CKA) \cite{Cristianini2006,kornblith2019} and RV coefficient \cite{Robert1976} by removing the biasing influence of measurement noise. 

The WUC and whitened RDM Pearson correlation have been implemented in a new Python-based RSA toolbox, released by the team of authors (\url{https://github.com/rsagroup/rsatoolbox}, \cite{Schuett2020}). We hope that the results presented in this paper, together with the accessible implementation, will accelerate the adoption of what we consider to be current best practice in RSA. 

\section{Methods}
\subsection{Extended definitions}
To derive the mean and full variance-covariance matrix of the distance estimates, it is useful to make some more general definitions and assumptions. We assume that each measured activity profile (column of $\hat{\mathbf{B}}_m$) has covariance  $\boldsymbol{\Sigma}_{K}$ between conditions. For fMRI, this correlation structure is caused by the sluggish nature of the hemodynamic response, as well from the low-frequency noise inherent to the measurements. A reasonable estimate of $\boldsymbol{\Sigma}_K$ can be derived directly from the first level linear model \cite{Friston1995}. For other modalities, it may be more reasonable to assume independence of measurements.

We also assume that each measured activity pattern (row of $\hat{\mathbf{B}}_m$) has a covariance of $\boldsymbol{\Sigma}_{P}$ between channels. Variability of fMRI, EEG, MEG measurements clearly shows substantial spatial structure. Again, this noise structure can usually be estimated from the residuals of the first-level linear model (see Mahalanobis distance). To remove the redundancy of $\boldsymbol{\Sigma}_{K}$ and $\boldsymbol{\Sigma}_{P}$ in terms of the overall scaling of the noise, we restrict $\mathrm{trace}(\boldsymbol{\Sigma}_{P})=P$.  

For the derivation of the full covariance matrix of distance estimates, we need to make the slightly more restrictive assumption that $\hat{\mathbf{B}}_m$ has a matrix-normal distribution across partitions $m$. While this assumption is reasonable for fMRI data, it is recommended to apply the square-root transform to neuronal spiking data to make it conform to the normal assumption \cite{Yu2009}.

To derive distances between conditions in a matrix notation, we define a $D \times K$ contrast matrix $\mathbf{C}$. The $k^{th}$ row of this matrix contains a $1$ and a $-1$ for the two conditions that are contrasted in the $k^{th}$ distance, all other entries are $0$. The product $\mathbf{CB}$ then results in a $D \times P$ matrix that contains the pattern differences $\boldsymbol{\delta}_k =\mathbf{b}_i - \mathbf{b}_j$ in its rows. We define:

\begin{equation}
    \boldsymbol{\Delta} \equiv  \mathbf{C}\mathbf{B}\,\mathbf{B}^T\mathbf{C}^T /P \label{eq:define_delta}
\end{equation}

The diagonal of $\boldsymbol{\Delta}$ contains the squared distances $d_k$ (divisively normalized by the number of channels). On the basis of $\boldsymbol{\Sigma}_K$, we can also define the $D \times D$ variance-covariance matrix of the pattern-difference estimates ($\mathbf{C}\hat{\mathbf{B}}_m$): 

\begin{equation}
    \boldsymbol{\Xi} \equiv \mathrm{Var}(\mathbf{C}\,\hat{\mathbf{B}}_m)=\mathbf{C}\boldsymbol{\Sigma}_K\mathbf{C}^T.
    \label{eq:define_Xi}
\end{equation}

\begin{table}[!bt]
\caption{\textbf{Table of symbols.}}
\begin{threeparttable}
\begin{tabular}{lll}
\headrow
\thead{Symbol} & \thead{Size} & \thead{Meaning}\\
\hline 
 $K$ & 1 & Number of conditions \\
 $P$ & 1 & Number of channels or voxels\\
 $M$ & 1 & Number of independent data partitions\\
 $D$ & 1 & Number of distances, usually $K(K-1)/2$\\
 $N_m$ & 1 & Number of activity measurements for partition $m$\\
 \hline 
 $\mathbf{B}$ & $K \times P$ & True activity patterns\\
 $\mathbf{b}_i$ & $1 \times P$ & True activity patterns for condition $i$\\
 $\hat{\mathbf{B}}_m$ & $K \times P$  & Estimated or measured activity patterns for partition $m$\\
$\boldsymbol{\Sigma}_K$ & $K \times K$  & Variance-covariance matrix of $\hat{\mathbf{B}}_m$ between conditions\\
$\boldsymbol{\Sigma}_P$ & $P \times P$  & Variance-covariance matrix of $\hat{\mathbf{B}}_m$ between channels or voxels\\
 \hline 
 $\boldsymbol{\delta}_k$ & $1 \times P$ & True pattern difference for condition pair $k$\\
  $d_k$ & $1$ & True distance for condition pair $k$ \\
 $\mathbf{d}$ & $D \times 1$ & Vector of all pairwise distances \\
 $\hat{\mathbf{d}}$ & $D \times 1$ & Biased distance estimates \\    
 $\tilde{\mathbf{d}}$ & $D \times 1$ & Unbiased distance estimates  \\
 $\boldsymbol{\Delta}$ & $D \times D$ & Matrix of inner products of all pattern differences\\
$\boldsymbol{\Xi}$ & $D \times D$ & Variance-covariance matrix of all estimated pattern differences\\
$\mathbf{V}$ & $D \times D$ & Variance-covariance matrix of distance estimates\\
\hline
\end{tabular}
\begin{tablenotes}
\item Size is given in number of rows $\times$ number of columns. For consistency of notation, vectors are defined to be in either row or column orientation.
\end{tablenotes}
\end{threeparttable}
\end{table}

\subsection{Bias of distance estimates}

Eq. \ref{eq:distbias} can be derived by expressing the estimated pattern difference ($\hat{\boldsymbol{\delta}}_{k,m}$) as the sum of the true pattern-difference vectors $\boldsymbol{\delta}_{k}$ and the measurement noise ($\boldsymbol{\xi}_{k,m}$). By substituting this into Eq. \ref{eq:biasedDist} and taking the expected value ($E$), it is straightforward to show that the distance estimator is positively biased: 

\begin{equation}
\begin{split}
\mathrm{E}(\hat{d_k})&=\mathrm{E}\left(\frac{1}{M^2}\sum_{m} \sum_{n} (\boldsymbol{\delta}_{k}+\boldsymbol{\xi}_{k,m})(\boldsymbol{\delta}_{k}+\boldsymbol{\xi}_{k,n})^T/P\right) \\
&=\left(\boldsymbol{\delta}_k\boldsymbol{\delta}_k^T + \mathrm{E} \left( \sum_{m}\boldsymbol{\xi}_{k,m} \boldsymbol{\xi}_{k,m} ^ T/M \right) \right)/P \\
&=d_k+\boldsymbol{\Xi}_{kk}/M.
\end{split}
\label{eq:distbias2}
\end{equation}

The bias arises by multiplying the noise with itself (i.e. for the cases of $m=n$). For all other cases ($m\neq n$), the noise terms are independent, and the expected value of their product is zero. 

\subsection{Variance of distance estimates}

An analytical expression for the variance-covariance matrix of the vector of distance estimates can be derived using the following general result (see Appendix B1, B2 for details). If the matrix $\mathbf{A}$ has matrix normal distribution $\mathcal{MN}(\mathbf{B},\boldsymbol{\Xi},\boldsymbol{\Sigma})$, then the diagonal of $\mathbf{A}\mathbf{A}^T$ has the expected value and variance: 

\begin{eqnarray}
\rm E(\rm diag(\mathbf{A}\mathbf{A}^T))&=&\rm diag(\mathbf{B}\mathbf{B}^T+\rm tr(\boldsymbol{\Sigma})\boldsymbol{\Xi})\\
\rm Var(\rm diag(\mathbf{A}\mathbf{A}^T))&=&4\mathbf{B}\boldsymbol{\Sigma}\mathbf{B}^T \circ \boldsymbol{\Xi}+2\rm tr(\boldsymbol{\Sigma\Sigma})(\boldsymbol{\Xi} \circ \boldsymbol{\Xi}),
\label{eq:VardiagMN}
\end{eqnarray}
where $\circ$ is the element-by-element multiplication of two matrices, and $\mathrm{tr}$() is the trace of a matrix. When setting $\mathbf{A}$ to the mean of the pattern differences across partitions, we can easily derive the variance of the biased distance estimate (Eq. \ref{eq:biasedDist}).

\begin{eqnarray}
\mathrm{Var}(\hat{\mathbf{d}})&=&\frac{1}{P^2}\left( \frac{2\mathrm{tr}(\boldsymbol{\Sigma}_P\boldsymbol{\Sigma}_P)}{M^2}\boldsymbol{\Xi}\circ \boldsymbol{\Xi} + \frac{4P}{M}\boldsymbol{\Delta}^*\circ \boldsymbol{\Xi} \right) \label{eq:cov_biased}\\
\boldsymbol{\Delta}^*&=&\mathbf{CB}\boldsymbol{\Sigma}_{P}\mathbf{B}^T\mathbf{C}^T
\end{eqnarray}   

The expression for the variance of the unbiased estimate of the distance (Eq. \ref{eq:unbiasedDist}) can be derived by taking the covariance of all pairs of partitions into account (see Appendix B3).   

\begin{equation}
\mathrm{Var}(\tilde{\mathbf{d}})=\frac{1}{P^2}\left(\frac{2\mathrm{tr}(\boldsymbol{\Sigma}_P\boldsymbol{\Sigma}_P)}{M(M-1)}\boldsymbol{\Xi}\circ \boldsymbol{\Xi} + \frac{4P}{M}\boldsymbol{\Delta}^*\circ \boldsymbol{\Xi}\right)
\label{eq:cov_unbiased}
\end{equation}

Intuitively the variances of distance estimates come from the product of signal and noise (averaged over $M$ partitions) and product of noise with noise (averaged over $M^2$ pairs of partitions for the biased distance estimate and over $M(M-1)$ pairs of partitions for the unbiased estimate). 

\subsection{Spatial pre-whitening and Mahalanobis distances}
In the result section, we focus on biased and unbiased estimates of the Euclidean distance. Previous work \cite{Walther2016}, however, demonstrates clearly that taking into account the spatial covariance structure of fMRI noise ($\boldsymbol{\Sigma}_P$) can increase the reliability of distance estimates. 

In the simplest case, we ignore the correlation between voxels and simply divide the activity estimates for each voxel by the square root of the diagonal elements of $\boldsymbol{\Sigma}_P$. This step already prevents noisy voxels to influence the distance estimate overly much. Additionally, we can use multivariate pre-whitening, i.e. post-multiplication of $\boldsymbol{\Sigma}_P^{-1/2}$. This step gives less weight to the information contained in two voxels that are highly correlated in their random variability than to information contained in two uncorrelated voxels.  Calculating Euclidean distances on multivariate pre-whitened data is equivalent to calculating a Mahalanobis distance. 

In practice, we do not have access to the voxel-by-voxel covariance matrix. However, we can use the residuals $\mathbf{R}_m$ from the first-level general linear model to derive an estimate,

\begin{equation}
    \hat{\boldsymbol{\Sigma}}_P=\frac{1}{M(N_m-K_m)}\sum_{m=1}^{M}\mathbf{R}_m^T\mathbf{R}_m,
\label{eqWhiteSigmaP}
\end{equation}

where $N_m$ is the number of observations, and $K_m$ is the number of regressors of interest per partition. Oftentimes, we have the case that $P>N$, which renders the estimate non-invertible. Even with $N>P$, it is usually prudent to regularize the estimate, as it stabilizes the distance estimates. A practical way of doing this is to shrink the estimate of the covariance matrix to a diagonal version of the raw sample estimate:

\begin{equation}
\tilde{\boldsymbol{\Sigma}}_P=h\,\text{diag}(\hat{\boldsymbol{\Sigma}}_P)+(1-h)\,\hat{\boldsymbol{\Sigma}}_P.
\end{equation}

The scalar h determines the amount of shrinkage, with 0 corresponding to no shrinkage and 1 to only using the diagonal of the matrix (univariate prewhitening). Estimation methods for the optimal shrinkage coefficient have been proposed \cite{Ledoit2004}, but in practice values in the range of $h=[0.2-0.4]$ perform well for fMRI data. The estimate is then used to obtain a spatially prewhitened versions of $\hat{\mathbf{B}}_m$:

\begin{equation}
\hat{\mathbf{B}^*}_m=\hat{\mathbf{B}}_m\tilde{\boldsymbol{\Sigma}}_P^{-1/2}
\end{equation}

Biased and unbiased estimates of the Mahalanobis distance can then be calculated via Eq. \ref{eq:distbias}, \ref{eq:unbiasedDist}, using $\hat{\mathbf{B}}^*_m$ instead of $\hat{\mathbf{B}}_m$. To obtain a full expression of the variance-covariance matrix of this distance, we need to know the mean and covariance of the pre-whitened data. If whitening would work perfectly, the data would be independent across voxels. However, given that we operate with an estimate of $\boldsymbol{\Sigma}_P$, this is not the case. Rather, the pre-whitened data will have matrix-normal distribution 

\begin{eqnarray}
    \hat{\mathbf{B^*}}_{m} &\sim\mathcal{MN} (\mathbf{B}\tilde{\boldsymbol{\Sigma}}_P^{-1/2},\boldsymbol{\Sigma}_{K},\boldsymbol{\Sigma}_{R})\\
\boldsymbol{\Sigma}_R&=\tilde{\boldsymbol{\Sigma}}_P^{-1/2}\boldsymbol{\Sigma}_P\tilde{\boldsymbol{\Sigma}}_P^{-1/2}
\end{eqnarray}

The covariance matrix of these distance estimates is given by Eq. \ref{eq:cov_biased}, \ref{eq:cov_unbiased}, with $\mathbf{B}$ replaced by $\mathbf{B}\tilde{\boldsymbol{\Sigma}}_P^{-1/2}$, and $\boldsymbol{\Sigma}_P$ with $\boldsymbol{\Sigma}_R$. The covariance structure under the assumption that $\mathbf{B}=\mathbf{0}$, however, will be the same as for the Euclidean distance - therefore the whitened RDM correlation and the WUC can be use equivalently for the biased and unbiased estimates of the Mahalanobis distance.   

\subsection{Simulations}

To evaluate different ways of comparing RDMs, we conducted a range of simulations, each from a known true model. Each model specified a second moment matrix. For every simulated data set, we first generated the matrix $\mathbf{B}$ of true activity patterns from a multivariate normal distribution, such that the second-moment matrix ($\mathbf{B}^T\mathbf{B}/P$) was exactly $\mathbf{G}s$, with $\mathbf{G}$ being the second moment matrix specified by the model and $s$ the scalar signal strength. The true patterns were independent across voxels. We then generated a set of $M$ measured activity patterns with 

$$
\hat{\mathbf{B}}_m=\mathbf{B} + \mathbf{E}_m,
$$

where the noise $\mathbf{E}_m$ was drawn from a matrix normal distribution with mean zero, column covariance $\boldsymbol{\Sigma}_P$ and row covariance $\boldsymbol{\Sigma}_K$. For most simulations, the noise was independent across channels $\boldsymbol{\Sigma}_P=\mathbf{I}_P$, and observations $\boldsymbol{\Sigma}_K=\mathbf{I}_K$. The influence of spatial noise correlations is studied in Appendix 6.3, and the influence of condition-wise correlations in the example in Figure 4a. 

Each simulated data set was compared to the different candidate models using the different criteria.  We then counted how often each method decided for the true (i.e., data generating) model over the competing model. 

For the results shown in Figure 4, we used 2 simple models, each specifying different dissimilarities across 4 conditions. Data sets were generated from each of the two models with 2-12 partitions. The variance of the noise of the simulation was set to be proportional to the number of partitions, such that the variance of the averaged activity patterns was always constant. The noise was assumed to be independent across the $P=50$ voxels. In simulation for Fig 4a, the measurement noise was correlated for neighboring conditions with $r=0.15$, for all other simulations it was independent. 

For the simulations shown in Figure 6, we used three example experiments from published fMRI studies. The first two examples come from a paper investigating the representational structure of finger movements in primary motor and sensory cortex \cite{Ejaz2015}. In Experiment 1, the activity patterns for K=5 fingers were measured. There were two candidate models, one that predicted the similarity structure based on the natural statistics of movement, the other based on the similarity of muscle activity patterns. The RDM correlation between the two models was relatively high ($r=0.85$). 

Experiment 2 (Experiment 3 in \cite{Ejaz2015}), compared  31 different finger movements, which span the whole
space of possible ``piano-chord'' combinations. Again, the predictions of the natural statistics and muscle activity model were compared. 

The third example uses an experiment investigating the response of the human inferior temporal cortex to 96 images, including animate and inanimate objects \cite{Khaligh-Razavi2014}. The model predictions
are derived from a convolutional deep neural network model -- with each of the 7 layers providing a separate representational model. The bitmap images were presented to the deep neural network and the internal activity patterns used as representational models.

All data for Figure 6 were simulated with 8 runs, 160 voxels, and independent noise on the observations. The noise variance was set to
$\sigma^{2} = 1$. We first normalized the model predictions, such that the norm of the predicted squared Euclidean distances was 1. We then varied the strength of the signal systematically from 0 (pure noise data) to a level that achieved reasonably high accuracy. 
We generated 3,000 data sets for each experiment, parameter setting, and model. For Experiment 3, where there were 7 alternative models, we generated data sets from each of the models. We then decided whether the data was better fit by the data-generating or one of the alternative models. Accuracy was then averaged over all possible model pairs. Thus, for all 3 Experiments, chance performance was at 0.5. The code for the simulations presented in this paper is available at \url{https://github.com/rsagroup/rsaModelComparison}.

\section{acknowledgements}
This study was supported by the Discovery Grant from the National Science and Engineering Research Council (NSERC RGPIN-2016-04890) to J.D., the NSERC Discovery Grant (NSERC RGPIN-2019-06742) to M.M., the Canada First Research Excellence Fund (BrainsCAN) to Western University, and the German Research Foundation (DFG SCHU 3351/1-1) to H.S.

\bibliography{references.bib}
\newpage

\section{Appendix}

\subsection{Derivation of variance-covariance matrix of the distance estimate}
\subsubsection{Expectations of products of random variables}
The variance of distance estimates can be derived from the basic expectations of products of random variables. The expected value of the product any pair of random variable $x,y$ 

\begin{equation}
    \mathrm{E}\left(x\cdot y\right)=\mathrm{E}(x)\cdot\mathrm{E}(y)+{\rm cov}(x,y)
\end{equation}

If $u,v,x,y$ are jointly normally distributed, we have the following general expectations: 

\begin{equation}
\begin{split}
{\rm cov}\left(xy,uv\right)&=\mathrm{E}(x)\mathrm{E}(u){\rm cov}(y,v)+\mathrm{E}(x)\mathrm{E}(v){\rm cov}(y,u)\\
&+\mathrm{E}(y)\mathrm{E}(u){\rm cov}(x,v)+\mathrm{E}(y)\mathrm{E}(v){\rm cov}(x,u)\\
&+{\rm cov}(x,u){\rm cov}(y,v)+{\rm cov}(x,v){\rm cov}(y,u)
\label{eq:normalExp}
\end{split}
\end{equation}

\subsubsection{Expectations of the product of normal matrices}

From Eq. \ref{eq:normalExp}, we can derive the basic expectations on the product of normal random matrices. Let us assume that vector $\mathbf{x}$ has multi-variate normal distribution with mean $\boldsymbol{\mu}$ and variance-covariance matrix $\mathbf{V}$. To derive Eq. \ref{eq:cov_biased} and \ref{eq:cov_unbiased}, we require the following results for the outer product $\mathbf{xx}^T$. The mean is given by  

\begin{equation}
\rm E(\mathbf{x}\mathbf{x}^T)=\boldsymbol{\mu}\boldsymbol{\mu}^T+\mathbf{V}.
\end{equation}

The variance-covariance matrix of the diagonal of $\mathbf{xx}^T$ is

\begin{equation}
\rm Var(\rm diag(\mathbf{xx}^T))=4\boldsymbol{\mu}\boldsymbol{\mu}^T \circ \mathbf{V}+2(\mathbf{V}\circ\mathbf{V})
\end{equation}

These results can be easily extended to the distribution of the matrix product $\mathbf{X}\mathbf{X}^T$, where $\mathbf{X}$ is a random $N \times P$ matrix with independent normally-distributed columns, i.e. with matrix normal distribution $\mathbf{X} \sim \mathcal{MN}(\mathbf{M},\mathbf{V},\mathbf{I})$. 

\begin{eqnarray}
\rm E(\mathbf{X}\mathbf{X}^T)&=&\mathbf{M}\mathbf{M}^T+P\mathbf{V}
\end{eqnarray}

The full variance-covariance matrix of the diagonal $\mathbf{d}$ of $\mathbf{X} \mathbf{X}^T$ is

\begin{equation}
\rm Var(\rm diag(\mathbf{XX}^T))=4\mathbf{M}\mathbf{M}^T \circ \mathbf{V}+2P(\mathbf{V}\circ\mathbf{V}).
\end{equation}

Finally, we need to generalize these results to a situation, where the columns of $\mathbf{X}$ are not independent, but have element-wise covariance of $\boldsymbol{\Sigma}$. Thus, we are interested in the joint distribution of the elements of the quadratic form $\mathbf{X}\mathbf{X}^T$, where $\mathbf{X}$ has matrix normal distribution $\mathbf{X} \sim \mathcal{MN}(\mathbf{M},\mathbf{V},\boldsymbol{\Sigma})$. 

\begin{eqnarray}
\rm E(\mathbf{X}\mathbf{X}^T)&=&\mathbf{M}\mathbf{M}^T+\rm tr(\boldsymbol{\Sigma})\mathbf{V}\\
\rm Var(\rm diag(\mathbf{XX}^T))&=&4\mathbf{M}\boldsymbol{\Sigma}\mathbf{M}^T \circ \mathbf{V}+2\rm tr(\boldsymbol{\Sigma\Sigma})(\mathbf{V}\circ\mathbf{V})
\label{eq:VardiagMN2}
\end{eqnarray}

From this result, we can obtain Eq. \ref{eq:cov_biased} by considering that the mean of $\mathbf{C}\hat{\mathbf{B}}_m$ across partitions has variance $\boldsymbol{\Xi}/M$. 

\subsubsection{Averaging across partitions}
To derive the variance of the unbiased distances, we need to take into account the averaging of the estimated difference across the $M$ different crossvalidation folds. While data from different partitions can be assumed to be independent, the inner products across crossvalidation folds are not. This is because the partitions from one crossvalidation fold will be again included in other folds. The two pattern differences that enter the product in Eq. \ref{eq:unbiasedDist} come from a single partition (that is, $A = {m}$), or from the set of all other partitions (that is, $B = M \setminus {m}$, which we will denote here in short by $\setminus {m}$. 

As a shorthand for the covariance between difference estimates $i$ and $j$ that are based on the set of partitions $A$ and $B$, we introduce the symbol

\begin{equation}
\label{eq:partCov}
    {\Xi}_{i,j}^{A,B}={\rm Cov}\big( \hat{\boldsymbol{\delta}}_{i,A},\,\hat{\boldsymbol{\delta}}_{j,B} \big).
\end{equation}

This is the covariance for each individual voxel. We now exploit the
bilinearity of the covariance operator, that is,

\begin{equation}
\label{eq:partCov2}
    {\rm Cov}\Big( \sum_{m} {x}_m , \, \sum_{n} {y}_n \Big)=\sum_{m} \sum_{n} {\rm Cov}\big( {x}_m,\,{y}_n \big),
\end{equation}

to obtain the following general result:

\begin{equation}
    \begin{split}
    {\rm Cov}(\tilde{d}_i,\,\tilde{d}_j)
    =& \frac{1}{{M^2P^2}} \sum_{m} \sum_{n} {\rm Cov}\big( \hat{\boldsymbol{\delta}}_{i,m}\hat{\boldsymbol{\delta}}_{i,\setminus m}^T,\, \hat{\boldsymbol{\delta}}_{j,n}\hat{\boldsymbol{\delta}}_{j,\setminus n}^T  \big) \\
    =&\frac{1}{M^2P^2} 
    \sum_{m} \sum_{n}  P \boldsymbol{\delta}_i \boldsymbol{\Sigma}_P \boldsymbol{\delta}_j^T \Big({\Xi}_{i,j}^{\setminus m\setminus n}+{\Xi}_{i,j}^{\setminus m n}+{\Xi}_{i,j}^{ m\setminus n}+{\Xi}_{i,j}^{ m n}\Big)   \\
    &+ \text{tr}\big(\boldsymbol{\Sigma}_{P}\boldsymbol{\Sigma}_P \big) \Big({\Xi}_{i,j}^{ m n}{\Xi}_{i,j}^{\setminus m\setminus n}+{\Xi}_{i,j}^{ m\setminus n}{\Xi}_{i,j}^{\setminus m n} \Big)\ \\
    =&\frac{1}{P^2} \Big\{ P \boldsymbol{\delta}_i \boldsymbol{\Sigma}_P  \boldsymbol{\delta}_j^T   {S}_{i,j}+\rm tr\big(\boldsymbol{\Sigma}_{P}\boldsymbol{\Sigma}_P\big){N}_{i,j} \Big\} 
\label{eq:partCov3}
\end{split}
\end{equation}

where

\begin{equation}
        {S}_{i,j}=\frac{1}{M^2}\sum_{m} \sum_{n}  \Big\{ {\Xi}_{i,j}^{\setminus m\setminus n}+{\Xi}_{i,j}^{\setminus m n}+{\Xi}_{i,j}^{ m\setminus n}+{\Xi}_{i,j}^{ m n} \Big\} 
\label{eq:partCovS}
\end{equation}

and

\begin{equation}
        {N}_{i,j}=\frac{1}{M^2}\sum_{m} \sum_{n}  \Big\{ {\Xi}_{i,j}^{ m n}{\Xi}_{i,j}^{\setminus m\setminus n}+{\Xi}_{i,j}^{ m\setminus n}{\Xi}_{i,j}^{\setminus m n} \Big\}.  
\label{eq:partCovN}
\end{equation}

This is the most general expression of the variance of the unbiased distance, which can even be used when the covariance structure of different partitions ($\boldsymbol{\Sigma}_K$) differs from each other (see Appendix C). 

For the case in which the difference estimates from all $M$ partitions can be assumed to have the same covariance, that is,
$\mathbf{\Xi}_{i,j}\equiv {\rm Cov}(\hat{\boldsymbol{\delta}}_{i,m},\,\hat{\boldsymbol{\delta}}_{j,m})$, we can simplify the expression dramatically. In this instance the best estimate of $\boldsymbol{\delta}_{\setminus m}$ is the average of all partitions except $m$:

\begin{equation}
    \hat{\boldsymbol{\delta}}_{\setminus m}=\sum_{n\neq m} \hat{\boldsymbol{\delta}}_n/(M-1).
    \label{eq:partCov4}
\end{equation}

Accordingly, for $n\neq m$, we have

\begin{equation}
\boldsymbol{\Xi}^{A,B} = \left\{
\begin{array}{lll}
\boldsymbol{\Xi} & \mbox{if }     A=m,\,B=m\\
\mathbf{0} &  \mbox{if }         A=m,\,B=n\\
\mathbf{0} &  \mbox{if }         A=m,\,B=\setminus m\\
\boldsymbol{\Xi}/(M-1) & \mbox{if } A=\setminus m,\,B=\setminus m\\
\boldsymbol{\Xi}/(M-1) & \mbox{if }     A=m,\,B=\setminus n\\
(M-2)\boldsymbol{\Xi}/(M-1)^2 & \mbox{if } A=\setminus m,\,B=\setminus n\\
\end{array}
\right.
\end{equation}.

Substituting the elements of the appropriate representations of
$\boldsymbol{\Xi}^{A,B}$ into Eq. \ref{eq:partCovS}, \ref{eq:partCovN} and summing up, we have

\begin{equation}
\begin{split}
    {S}_{i,j}=\ &\frac{1}{M^2}\Big\{ M\big( \frac{{\Xi}_{i,j}}{M-1} + {\Xi}_{i,j} \big) \\
    &+M(M-1) \Big( \frac{(M-2){\Xi}_{i,j}}{(M-1)^2}+\frac{2{\Xi}_{i,j}}{M-1} \Big)\,\, \Big\} \\
    =\ &\frac{1}{M^2}\,{\Xi}_{i,j} \Big\{ \frac{M}{M-1}+M+\frac{M(M-2)}{M-1}+2M \Big\} \\
    =\ &\frac{4}{M}\,{\Xi}_{i,j}, \\
    {N}_{i,j}=\ &\Big\{ \frac{M{\Xi}_{i,j}{\Xi}_{i,j}}{M-1}+\frac{M(M-1){\Xi}_{i,j}{\Xi}_{i,j}}{(M-1)^2}  \Big\}  \\
    =\ &\frac{2\,{\Xi}_{i,j}{\Xi}_{i,j}}{M(M-1)}, 
\end{split}
\label{eqAvrg7}
\end{equation}

and

\begin{equation}
        {\rm Cov}(\tilde{d}_i,\,\tilde{d}_j)=\frac{1}{P^2} 
        \Big( 4\frac{P\boldsymbol{\delta}_{i}\boldsymbol{\Sigma}_P\boldsymbol{\delta}_{j}^{T}}{M}
        \Xi_{i,j}+\frac{2\rm tr\big(\boldsymbol{\Sigma}_{P}\boldsymbol{\Sigma}_P \big)}{M(M-1)}\Xi_{i,j}\Xi_{i,j} \Big). 
\end{equation}

Finally, on writing the desired complete covariance matrix using
element-by-element multiplication (Hadamard product), we obtain the result given in
Eq. \ref{eq:cov_unbiased}.

\subsection{Unbalanced designs}

In unbalanced designs, the noise covariance $\boldsymbol{\Sigma}_K$ is different across different partitions - i.e. each partition has their own covariance matrix $\boldsymbol{\Sigma}_{K}^m$. In the calculation of the distances, this ideally should be taken into account. To simplify this problem we here assume that all signals are zero, as we did for the derivation of the WUC. 

Given the zero mean assumption and independent runs, the product of patterns from one pair of partitions, is uncorrelated to the product of two patterns from any other pair of partitions (Eq.\ref{eq:normalExp}). Thus, for both biased and unbiased distance estimates, the optimal pooling of the estimates from the pairs is their precision weighted average.

The distance estimate from single pair of partitions $\hat{\mathbf{d}}_{m,n}=\mathrm{diag}(\mathbf{C}\hat{\mathbf{B}}_m\hat{\mathbf{B}}_n^T\mathbf{C})/P$ has the following expected value ($\mathrm{E}$) and covariance ($\mathrm{Var}$):

If $m\neq n$:
\begin{equation}
    \mathrm{E}(\hat{\mathbf{d}}_{m,n}) = 0
\end{equation}

\begin{equation}
\mathrm{Var}(\hat{\mathbf{d}}_{m,n}) = \mathrm{tr}(\boldsymbol{\Sigma_P}\boldsymbol{\Sigma_P}) \boldsymbol{\Xi}^m \circ \boldsymbol{\Xi}^n = \mathrm{tr}(\boldsymbol{\Sigma_P}\boldsymbol{\Sigma_P})(\mathbf{C} \boldsymbol{\Sigma}_k^m \mathbf{C}^T) \circ (\mathbf{C} \boldsymbol{\Sigma}_k^n \mathbf{C}^T)
\end{equation}

If $m = n$:
\begin{equation}
    \mathrm{E}(\hat{\mathbf{d}}_{m,m}) = \mathrm{tr}(\boldsymbol{\Sigma_P}\boldsymbol{\Sigma_P}) \mathrm{diag}(\mathbf{C} \boldsymbol{\Sigma}_k^m \mathbf{C}^T)
\end{equation}
\begin{equation}
\mathrm{Var}(\hat{\mathbf{d}}_{m, m}) = \mathrm{tr}(\boldsymbol{\Sigma_P}\boldsymbol{\Sigma_P}) \boldsymbol{\Xi}^m \circ \boldsymbol{\Xi}^n = \mathrm{tr}(\boldsymbol{\Sigma}_P\boldsymbol{\Sigma}_P) (\mathbf{C} \boldsymbol{\Sigma}_k^m \mathbf{C}^T) \circ (\mathbf{C} \boldsymbol{\Sigma}_k^m \mathbf{C}^T)
\end{equation}

Using these formulas we derive an optimal unbiased estimate using precision weighting:
\begin{equation}
    \tilde{\mathbf{d}} = \left(\sum_m \sum_{n \neq m} \mathrm{Var}^{-1}(
    \hat{\mathbf{d}}_{m,n}) \right) ^{-1} \sum_m \sum_{n \neq m} \mathrm{Var}^{-1}(\hat{\mathbf{d}}_{m,n}) \hat{\mathbf{d}}_{m,n}
\end{equation}

The covariance matrix of this combined estimate is then:

\begin{equation}
    \left(\sum_m \sum_{n \neq m} \mathrm{Var}^{-1}(
    \hat{\mathbf{d}}^{m,n}) \right) ^{-1}
\end{equation}

\subsection{Simulations with spatially correlated noise}
To study the influence of spatial noise correlations, we repeated the simulations for Experiment 2 (see Fig. 6), using noise that was spatially correlated across voxels,  $\Sigma_P \neq \mathbf{I}$. To parametrically vary the level of spatial co-dependence, we generated the data on a $6\times6\times6$ cube of isotropic voxels. The covariance between the noise term of the two voxels $e_i$ and $e_j$ was determined by

$$
\mathrm{cov}(e_i,e_j)=\mathrm{exp}(-d_{i,j}^2/s^2) 
$$

with $d_{i,j}$ being the distance between two voxels (in voxel widths), and $s^2$ the variance of the Gaussian covariance kernel, which ranged from $0$ (independent noise) to $5$ (highly correlated noise). The "effective number" of voxels,  $\mathrm{tr}(\boldsymbol{\Sigma}_P)^2/\mathrm{tr}(\boldsymbol{\Sigma}_P\boldsymbol{\Sigma}_P)$,  was $216$ for $s^2=0$ and $15$ for $s^2=5$. The signal patterns ($\mathbf{B}$) for the 31 conditions were generated to be spatially independent, with a row-covariance as specified by either of the 2 candidate models.  

As expected, the accuracy of model selection decreased as the spatial correlation in the noise increased. Nonetheless, the advantage of the whitened RDM Pearson correlation and whitened unbiased cosine similarity over their traditional non-whitened cousins remained stable . 

Interestingly, PCM was superior to the whitened RDM correlation only for spatially independent, or for nearly independent, data. For a high degree of spatial covariance ($s^2=5$), the model selection accuracy was higher for whitened Pearson correlation (64.3\%) than for PCM (62.4\%). Thus, the likelihood-based test implemented in PCM is only better for model comparison for data where the residual covariances are limited to a size that typically remain after multivariate prewhitening of fMRI data \citep{Diedrichsen2016ArXiv}. For highly correlated raw fMRI data (non- or uni-variately prewhitented), RSA-based methods can perform slightly superior. 

\begin{figure}[h!]
\centering
\includegraphics[width=0.5\textwidth]{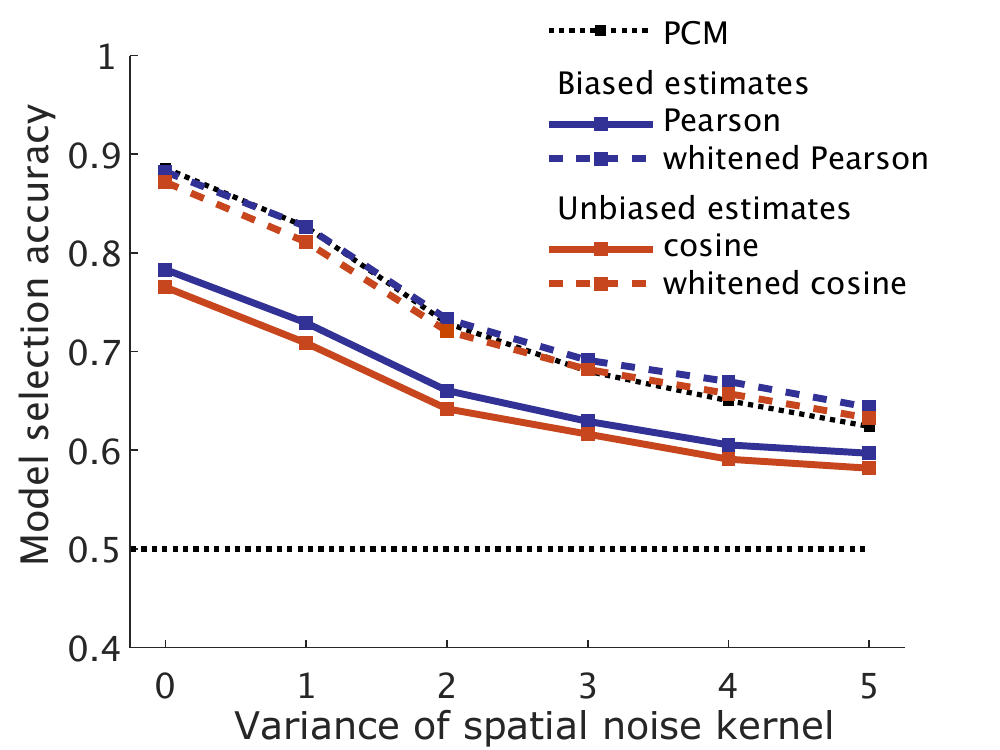}
\caption{\textbf{Simulation of Experiment 2 with spatially correlated noise}. Model selection accuracy for data generated for $6^3=216$ voxel cube with noise that had a Gaussian smoothness of 0-5.}
\end{figure}
\end{document}